\documentclass[acmsmall]{acmart}

\usepackage[ruled,vlined]{algorithm2e}
\usepackage{algorithmic}
\usepackage{amsmath}
\usepackage{bm}
\usepackage{multirow}
\usepackage{tabularx}
\usepackage{graphicx}
\usepackage{subfigure}
\newtheorem{Theorem}{Theorem}
\newtheorem{Definition}{Definition}

\AtBeginDocument{%
  \providecommand\BibTeX{{%
    \normalfont B\kern-0.5em{\scshape i\kern-0.25em b}\kern-0.8em\TeX}}}

\setcopyright{acmcopyright}
\copyrightyear{2018}
\acmYear{2018}
\acmDOI{10.1145/1122445.1122456}

\acmJournal{TOIS}
\acmVolume{37}
\acmNumber{4}
\acmArticle{111}
\acmMonth{8}

\begin{document}

\title{Reinforcement Routing on Proximity Graph for Efficient Recommendation}

\author{Chao Feng}
\author{Defu Lian}
\email{chaofeng@mail.ustc.edu.cn}
\affiliation{%
  \institution{University of Science and Technology of China}
  \streetaddress{No.443 Huangshan Road}
  \city{Hefei}
  \country{China}
  \postcode{230022}
}


\author{Xiting Wang}
\author{Zheng Liu}
\author{Xing Xie}
\affiliation{%
  \institution{Microsoft Research Asia}
  \streetaddress{No.5 Danleng Street}
  \city{Beijing}
  \country{China}
  \postcode{230022}}
\email{xitwan@microsoft.com}
\email{zhengliu@microsoft.com}
\email{xing.xie@microsoft.com}


\author{Enhong Chen}
\affiliation{%
  \institution{University of Science and Technology of China}
  \streetaddress{No.443 Huangshan Road}
  \city{Hefei}
  \country{China}
  \postcode{230022}}
\email{cheneh@ustc.edu.cn}
\authorsaddresses{
Authors’ addresses: Chao Feng, Defu Lian (corresponding author) and Enhong Chen, School of Computer Science and Technology, University of Science and Technology of China, No.443 Huangshan Road, Hefei, China, chaofeng@mail.ustc.edu.cn, \{liandefu, cheneh\}@ustc.edu.cn; Xiting Wang, Zheng Liu and Xing Xie, Microsoft Research Asia, No.5 Danleng Street, Beijing, China, \{xitwan,zhengliu,xing.xie\}@microsoft.com}
\renewcommand{\shortauthors}{Feng and Lian, et al.}

\begin{abstract}


We focus on Maximum Inner Product Search (MIPS), which is an essential problem in many machine learning communities. 
Given a query, MIPS finds the most similar items with the maximum inner products. Methods for Nearest Neighbor Search (NNS) which is usually defined on metric space don't exhibit the satisfactory performance for MIPS problem since inner product is a non-metric function. However, inner products exhibit many good properties compared with metric functions, such as avoiding vanishing and exploding gradients. As a result, inner product is widely used in many 
recommendation systems, which makes efficient Maximum Inner Product Search a key for speeding up many recommendation systems.

Graph based methods for NNS problem show the superiorities compared with other class methods. Each data point of the database is mapped to a node of the proximity graph. Nearest neighbor search in the database can be converted to route on the proximity graph to find the nearest neighbor for the query. This technique can be used to solve MIPS problem. Instead of searching the nearest neighbor for the query, we search the item with maximum inner product with query on the proximity graph. In this paper, we propose a reinforcement model to train an agent to search on the proximity graph automatically for MIPS problem if we lack the ground truths of training queries. If we know the ground truths of some training queries, our model can also utilize these ground truths by imitation learning to improve the agent's search ability. By experiments, we can see that our proposed mode which combines reinforcement learning with imitation learning shows the superiorities over the state-of-the-art methods 

\end{abstract}

\ccsdesc[500]{Information systems~Retrieval models and ranking}
\ccsdesc[500]{Information systems~Similarity measures}
\ccsdesc[500]{Information systems~Learning to rank}
\ccsdesc[500]{Information systems~Top-k retrieval in databases}

\keywords{MIPS, Non-metric, Proximity graph, Reinforcement learning, Reward shaping, Graph Convolutional Network.}

\maketitle

\section{Introduction}
In recent decades, the Maximum Inner Product Search (MIPS) in high dimensional space is a fundamental problem which has become a popular paradigm for solving large scale classification and retrieval tasks. It's wide application contains matrix factorization based recommendation systems \cite{10.1145/2396761.2396831,5197422,10.1145/3035918.3064009,10.5555/2976040.2976207,xue2017deep,8890891,8950031,10.1145/3366423.3380151,10.1145/3366423.3380187}, multi-class label prediction \cite{6619081,5206651}, natural language processing \cite{bengio2003neural,vaswani2017attention} and memory networks training \cite{chandar2016hierarchical}. For example, in matrix factorization based recommendation systems, users and items are embedded into dense vectors with the same dimensionality by factorizing the user-item interaction matrix. Usually, user vectors are regarded as queries and item vectors are regarded as database vectors respectively. The higher inner products between queries and database vectors indicate the corresponding users' more interest to the corresponding items. Formally,
the MIPS problem can be formulated as follows. Given a large database $X=\{\bm{x}_1,\bm{x}_2,\dots,\bm{x_}n\}$ ($\bm{x}_i\in \mathbb{R}^d, \ i\in \{1,2,\dots,n\}$) and a query vector $\bm{q}\in \mathbb{R}^d$, we need to find a vector $\bm{x}^*\in X$ such that maximizing the inner product with query $\bm{q}$, i.e.
\begin{equation}\label{definition:mips}
\begin{aligned}
\bm{x}^*=\arg \max_{\bm{x}\in X}\langle \bm{q},\bm{x} \rangle
\end{aligned}
\end{equation}
where $\langle \bm{a},\bm{b}\rangle=\sum_{i=1}^d a_i\cdot b_i,\ \bm{a},\bm{b}\in \mathbb{R}^d$ and $a_i,b_i$ are the corresponding $i$-th entries. At some situations, not just the top $1$  but the top $k$ ($k>1$) database vectors that maximizing the inner products with query $\bm{q}$ are required. The more general problem requires to find a subset $\bar{X}=\{\bm{x}_{i_1}, \bm{x}_{i_2},\cdots, \bm{x}_{i_k}\}$ with size $k$ ($k\ge 1$) from $X$ such that
\begin{equation}\label{definition:k_mips}
\begin{aligned}
\min_{\bm{x}\in \bar{X}}\langle \bm{q},\bm{x} \rangle \ge \max_{\bm{x'}\in X\setminus \bar{X}} \langle \bm{q},\bm{x'} \rangle .
\end{aligned}
\end{equation}

A naive approach for MIPS problem is to compute all the inner products among query $\bm{q}$ and the vectors $\bm{x}\in X$ exhaustively. Then we can return the exact top $k$ solutions. However, the time complexity of exhaustive search is linear with the the scale of database. In real application, the scale of database can be tens of millions and even larger but tasks usually require immediate response. For example, in matrix factorization based recommendation system where inner product is used as matching score between user vector and item vector, one user queries the recommendation system and hopes to obtain the recommended items immediately so that she/he can chooses some necessary items quickly. Obviously, the exhaustive search can't satisfy the efficiency and how to solve the MIPS problem efficiently becomes an important task.

MIPS problem is similar to Nearest Neighbor Search (NNS) problem which appears in many tasks, such as topic modeling ~\cite{DBLP:journals/ml/LiWOW21} and image retrieval ~\cite{DBLP:conf/wacv/LiuRR07}. Given a large database $X=\{x_{1},x_2,\dots,x_{n}\}$($\bm{x}_i\in \mathbb{R}^d, \ i\in \{1,2,\dots,n\}$) and the query vector $\bm{q}\in \mathbb{R}^d$, NNS problem requires to find a subset $\bar{X}$ with size $k$($k\ge 1$) from $X$ such that  
\begin{equation}\label{definition:anns}
\begin{aligned}
\max_{\bm{x}\in \bar{X}}dist(\bm{q},\bm{x}) \le \min_{\bm{x'}\in X\setminus \bar{X}} dist (\bm{q},\bm{x'}) .
\end{aligned}
\end{equation}
where $dist(\bm{a},\bm{b})$ measures the distance between vector $\bm{a}$ and vector $\bm{b}$. In NNS problem, the measure function $dist(\cdot,\cdot)$ (e.g. the Euclidian distance, $dist(\bm{a},\bm{b})=||\bm{a}-\bm{b}||)$) is usually defined on metric space. The measure function defined on metric space deserves three nice properties: 1). Identify of indiscernibles (i.e. $dist(\bm{a},\bm{b})=0\Longleftrightarrow \bm{a}=\bm{b}$); 2). Symmetry (i.e. $dist(\bm{a},\bm{b})=dist(\bm{b},\bm{a})$); 3). Triangle inequality (i.e. $dist(\bm{a},\bm{c})\le dist(\bm{a},\bm{b})+dist(\bm{b},\bm{c})$). However, inner product is defined on non-metric space and it doesn't satisfy the identify of indiscernibles and the triangle inequality. Those well designed methods for NNS problem don't have performance guarantee for MIPS problem. Recently, some literature tries to reduce MIPS problem to traditional NNS problem like \cite{friedman1977algorithm,neyshabur2015symmetric}. They expand the database vectors and query vectors to higher dimensional space. After the data transformation, maximizing the inner product can be converted to minimize the Euclidian distance among the expanded query vectors and database vectors. Then the methods for NNS problem such as Locality-Sensitive hash \cite{neyshabur2015symmetric} and partition trees \cite{keivani2018improved} can be applied to solve the converted problem.

\begin{figure}[t!]
\subfigure{
\begin{minipage}[c]{0.30\linewidth}
\centering
    \includegraphics[width=1\linewidth]{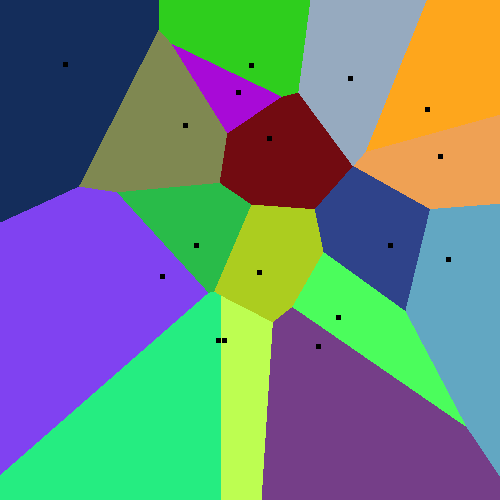}
\end{minipage}}
\ \ \ \ \ \ \ \
\subfigure{
\begin{minipage}[c]{0.30\linewidth}
\centering
    \includegraphics[width=1\linewidth]{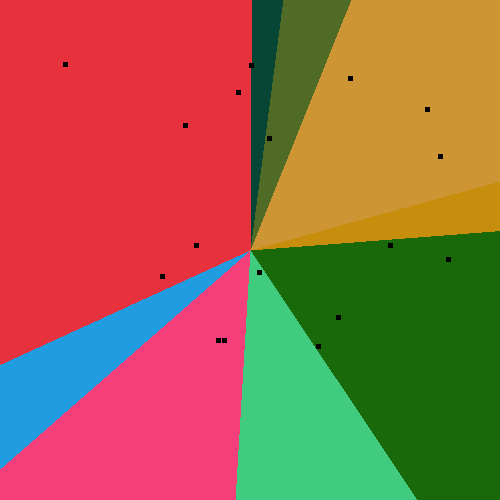}
\end{minipage}}
\caption{\textbf{Left:} The Voronoi Cell, $dist(\bm{a},\bm{b})=||\bm{a}-\bm{b}||$. \textbf{Right:} The s-Voronoi Cell, $s(\bm{a},\bm{b})=\langle \bm{a},\bm{b} \rangle$. }\label{fig:voronoi_cell}
\end{figure}

The significant superiorities of graph based methods for NNS problem also encourage researchers to solve MIPS problem by routing on the proximity graph \cite{morozov2018non,zhou2019mobius,tan2019efficient}. The database is mapped to a graph such that each data point in the database is represented by a vertex on the graph and the relevance among data points is exhibited by the edges. The NNS or MIPS problem becomes searching on the graph to find the candidate vertices. The key concept for graph based methods for NNS or MIPS problem is the Delaunay graph which is the dual graph of Voronoi diagram. For a data point $\bm{x}_i\in X$, the corresponding Voronoi Cell defines the region such that for any point in this region, it's nearest point in $X$ is $\bm{x}_i$. The formal definition is in \textsc{Definition \ref{vorono_cell}}.
\begin{Definition}[Voronoi Cell]{\label{vorono_cell}}
Given the database $X=\{\bm{x}_1,\bm{x}_2\cdots,\bm{x}_n\}$ ($\bm{x}_i\in \mathbb{R}^d, 1\le i\le n$) and a distance function $dist$ defined on metric space. The Voronoi Cell w.r.t. the point $\bm{x}_i$ is
\begin{displaymath}
R_i=\{\bm{x}\in \mathbb{R}^d |dist(\bm{x}_i,\bm{x})\le dist(\bm{x}_j,\bm{x}), \forall \bm{x}_j\in X\}.
\end{displaymath}
\end{Definition}
\begin{Definition}[Delaunay Graph]{\label{delaunay_graph}}
Given the database $X=\{\bm{x}_1,\bm{x}_2\cdots,\bm{x}_n\}$ ($\bm{x}_i\in \mathbb{R}^d, 1\le i\le n$) and a distance function $dist$ defined on metric space. The Delaunay Graph is a graph whose each node is corresponding to a data point in $X$ and there is a edge between node $i$ and node $j$ if their Voronoi Cells are adjacent (i.e. $R_i\cap R_j\neq \Phi$).
\end{Definition}
Intuitively, for any point $\bm{x}_i, \bm{x}_j\in X$, there is an edge between the corresponding node $i$ and node $j$ if and only if their Voronoi Cells share a common boundary. The distance function used in Voronoi Cell and Delaunay is defined on metric space which is not applicable for non-metric measure functions, such as inner product , the complicated item similarity model ~\cite{10.1145/3411754} in recommendation system and the cross-modal similarity ~\cite{10.1145/3408317,10.1145/3331184.3331196}. However, there are imitated definitions \cite{morozov2018non} for similarity functions defined on non-metric space as follows.
\begin{Definition}[s-Voronoi Cell \cite{morozov2018non}]{\label{s_vorono_cell}}
Given the database $X=\{\bm{x}_1,\bm{x}_2\cdots,\bm{x}_n\}$ ($\bm{x}_i\in \mathbb{R}^d, 1\le i\le n$) and a similarity function $s$ defined on non-metric space. The s-Voronoi Cell w.r.t. the point $\bm{x}_i$ is
\begin{displaymath}
R'_i=\{\bm{x}\in \mathbb{R}^d |s(\bm{x}_i,\bm{x})\ge s(\bm{x}_j,\bm{x}), \forall \bm{x}_j\in X\}.
\end{displaymath}
\end{Definition}
\begin{Definition}[s-Delaunay Graph \cite{morozov2018non}]{\label{s_delaunay_graph}}
Given the database $X=\{\bm{x}_1,\bm{x}_2\cdots,\bm{x}_n\}$ ($\bm{x}_i\in \mathbb{R}^d, 1\le i\le n$) and a similarity function $s$ defined on non-metric space. The s-Delaunay Graph is a graph whose each node is corresponding to a data point in $X$ and there is a edge between node $i$ and node $j$ if their s-Voronoi Cells are adjacent (i.e. $R'_i\cap R'_j\neq \Phi$).
\end{Definition}
 Besides the distance function $dist$ and similarity function $s$, the main difference between \textsc{Definition} \ref{vorono_cell} and \textsc{Definition} \ref{s_vorono_cell} is the inequality notation, i.e. "$\le$" for Voronoi Cell but "$\ge$" for s-Voronoi Cell.  \textbf{Figure \ref{fig:voronoi_cell}} exhibits the Voronoi Cell  of $dist(\bm{a},\bm{b})=||\bm{a}-\bm{b}||$ and s-Vononoi Cell of $s(\bm{a},\bm{b})=\langle \bm{a},\bm{b} \rangle$ respectively for dimensionality $d=2$. 
 
 Greedy search on Delaunay graph guarantees to find global optimal solutions for NNS problem  and MIPS problem ~\cite{malkov2014approximate,morozov2018non}. For database with size $n$ and dimensionality $d$, the time complexity is $\Omega(n^{\lceil d/2 \rceil})$ to construct exact Delaunay Graph (s-Delaunay Graph)~\cite{Th1987Preparata}. This time cost prohibits us to construct exact Delaunay graph for the database with large scale and high dimensionality. In addition, the Delauny graph reduces to the complete graph quickly as $d$ increases which is useless for efficient retrieval on the graph ~\cite{796589}. Usually, researchers compromise to construct a sparse and approximate Delaunay Graph (s-Delaunay Graph) for database with large scale and high dimensionality to solve NNS problem and MIPS problem.
In the later content, we also call the approximate Delaunay (s-Delauny) graph as proximity graph. 
The classical search methods based on greedy nature just explore the local structure of the graph and may make myopic decisions when searching on the graph. The literature \cite{baranchuk2019learning} reveals that the lack of graph's global information leads to being stuck in local optimum and the Graph Convolutional Network(GCN) can be used to relieve the vision limitation about the graph's global structure.  The authors of ~\cite{baranchuk2019learning} propose the imitation learning to route on the proximity graph which requires the ground truths (i.e. the solution of \textbf{E.q. (\ref{definition:mips})}.) of all the training queries. However, computing the ground truths for all training queries is infeasible for large-scale database which limits the method's usage in large-scale recommendation systems and retrieval systems. To consider all the above issues, we propose a new approach to solve MIPS problem. Our approach trains an agent by reinforcement learning to route on the proximity graph automatically. Graph Convolutional Network(GCN) is used to help the agent sense the global structure of the proximity graph.
Labeled data usually provides more information. Although calculating the ground truths for all training queries is time-consuming but we can calculate the ground truths for a few training queries. If the ground truths for some training queries are available, we utilize these ground truths to guide the agent. Concretely, we obtain the shortest path from location of the agent to the ground truth by existed algorithms(e.g. Dijkstra algorithm). Then the reward shaping is used to adjust the reward, i.e. we give a positive feedback if the agent walks along the shortest path and negative feedback if it doesn't follow the shortest path. Our approach can utilize both the labeled data (i.e. with ground truths) and the unlabeled data (i.e. without ground truths) by combining the reinforcement learning with imitation learning. We use the reinforce algorithm with self-critic baseline to update the routing policy of the agent. The experiment results verify the effectiveness of the proposed approach. As far as we know, this is the first work to utilize reinforcement learning with the help of imitation learning to tackle MIPS problem. We summarize our contributions as follows.

1.	To overcome the shortage of labeled data (i.e. we don't know the ground truths of the training queries), we firstly propose the reinforcement based method to train an agent to route on the proximity graph for MIPS problem. The Graph Convolutional Network (GCN) is utilized to help the agent sense the global structure of the proximity graph. 

2. If some ground truths for a few queries can be obtained by costing tolerable computation resources, we propose the composed model which combines reinforcement learning with imitation learning by reward shaping. The composed model utilizes both the labeled and unlabeled data to improve the performance of the agent.

3.	The reinforce algorithm with self-critical baseline is applied to train the proposed model which leads to faster convergence and lower variance.

4.	We extensively evaluate the proposed approach on three public datasets compared with the state-of-the-art methods. The experimental results show that the newly proposed method is consistently superior over the compared methods on all datasets.

The rest of the paper is organized as follows: We discuss related work in section 2. In section 3, we introduce the model used in our work. In section 4, we elaborate how to train the proposed model. In section 5, we introduce a few graph construction algorithms. Then, we present the empirical studies in section 6. Finally, we conclude our paper in section 7.

\section{Related work}
The methods for MIPS problem mainly fall into four groups: tree based methods, locality-sensitive hash based methods, quantization based methods and graph based methods.
KD-tree\cite{bentley1975multidimensional,bentley1979multidimensional}, RP-tree \cite{10.1007/s00453-014-9885-5} and Ball-tree \cite{omohundro1989five,2015arXiv151100628D} are widely used for NNS problem. Their idea is to partition the large dataset to many small datasets from coarse grain to fine grain by the tree structure, i.e. each child-node is a subset of it's parent node and the union of all nodes at each level makes up the whole dataset. Each leaf node is a small subset of the dataset and all items in the same leaf node have small distances with each other. Users can search the tree from root to leaf nodes and check all the items in chosen leaf nodes for a query. However, these tree based methods for NNS problem can't be used to MIPS problem directly as inner product is defined on non-metric space. For MIPS problem, Ram and Gray \cite{ram2012maximum} propose Cone tree. Compared with Ball tree, Cone tree replaces Euclidian distance with the angle of the two vectors as the measure function to split a set into two subsets. Cone tree requires that the angle of the two vectors which are chosen from different child nodes is large. Metric tree \cite{koenigstein2012efficient} is also constructed for MIPS problem. It uses a hyper-plane to partition points of a set into two subsets. The depth-first branch-and-bound algorithm is used to search on both the Cone tree and Metric tree. However, Shrivastava and Li \cite{NIPS2014_310ce61c} claim that tree based methods suffer the curse of dimensionality and the performance is weaker compared with other technique based methods. 

The locality-sensitive hash (LSH) based methods are independent of dimensionality of data point and get a lot of investigations for MIPS problem. LSH maps each data point into certain bucket and the similar items derive the higher probabilities to be mapped into the same bucket. The query is also mapped to a bucket and all the items in such bucket will be checked to find the top $k$ items for the query. LSH methods usually expand the vectors to higher dimensionality so that maximizing the original inner products can be converted to minimize the expanded vectors' Euclidian distances. Shrivastava and Li \cite{NIPS2014_310ce61c} argue that there is no symmetric LSH over the entire space $\mathbb{R}^d$. They propose the asymmetric LSH (referred as L2-ALSH) over the normalized queries (i.e. $||\bm{q}||=1$) and the dataset in unit sphere (i.e. $||\bm{x}||\le 1$, $\bm{x}\in X$) with $L_2$ hash function $h_{\bm{a},b}(\bm{x})=\lfloor \frac{\bm{a}^T\bm{x}+b}{r}\rfloor$. They expand the data point $\bm{x}$ by $P(\bm{x})=[U\bm{x};||U\bm{x}||^2;||U\bm{x}||^4;\cdots;||U\bm{x}||^{2^m}]^T$ and expand the query by $Q(q)=[\bm{q};1/2;\cdots;1/2]^T$. $\bm{a}\sim\mathcal{N}(0,I),b\sim\mathcal{U}(0,r)$. $m, U$ and $r$ are tunable parameters. They also propose the improved version, Sigh-ALSH \cite{10.5555/3020847.3020931}, by setting $P(\bm{x})=[\bm{x};1/2-||\bm{x}||^2;1/2-||\bm{x}||^4;\cdots;1/2-||\bm{x}||^{2^m}]^T$ and $Q(\bm{q})=[\bm{q};0;0;\cdots,0]^T$ with sign random projection hash function $h_{\bm{a}}(\bm{x})=sign(\bm{a}^T\bm{x})$. In the experiments, they argue that the LSH based methods show significant superiorities over tree based methods.
Neyshabur and Srebro \cite{10.5555/3045118.3045323} propose a parameter-free symmetric LSH (referred as Simple-LSH) over normalized queries and data points in unit sphere. They expand both the queries and data points by $P(\bm{x})=[\bm{x};\sqrt{1-||\bm{x}||^2}]^T$ with sign random projection hash function $h_{\bm{a}}(\bm{x})=sign(\bm{a}^T\bm{x})$. Similarly, Xbox method \cite{10.1145/2645710.2645741} expands the data point by $P(\bm{x})=[\bm{x};\sqrt{M^2-||\bm{x}||^2}]^T$ and expands the query by $Q(\bm{q})=[\bm{q};0]^T$ to avoid the transformation error, where $M$ is the maximum norm of all data points. Further more, H2-ALSH \cite{huang2018accurate} can avoid the transformation error so that maximizing the original vectors' inner product is exactly equivalent to minimizing the Euclidian distance of the expanded vectors. Compared with XBox method, the difference is that H2-ALSH uses $Q(\bm{q})=[\lambda \bm{q};0]^T$ ($\lambda=\frac{M}{||\bm{q}||}$) which leads to less distortion error.

The quantization technique traces back to signal processing \cite{gray1984vector,gersho2012vector}. Some classical methods like ITQ \cite{6296665}, OPQ \cite{ge2013optimized} and the improved version LOPQ \cite{kalantidis2014locally} are proposed for NNS problem. Quantization based methods approximate each data point by certain codeword of a codebook. The distances (similarities) between query and codewords make up the look-up table with much smaller scale compared with the scale of the dataset. Calculating the approximate  distance (similarity) between query and any data point by the look-up table is efficient. For MIPS problem, Du and Wang \cite{du2014inner} propose to use 
combination codes to quantize the data points. Concretely, for a data point $\bm{x}$ and a dictionary $\bm{C}\in \mathbb{R}^{d\times K}$, $M$ columns (denoted as $\bm{c}_{1}(\bm{x}),\bm{c}_{2}(\bm{x}),\cdots,\bm{c}_{M}(\bm{x})$) of $\bm{C}$ that minimize $||\bm{x}-\sum_{i=1}^M\bm{c}_{i}(\bm{x})||_2^2$ are selected. Then the sum of each codeword (i.e. $\sum_{i=1}^M\bm{c}_{i}(\bm{x})$) is the approximation of $\bm{x}$. If $\bm{c}_{1}(\bm{x}),\bm{c}_{2}(\bm{x}),\cdots,\bm{c}_{M}(\bm{x})$ are selected from $M$ different dictionaries $\bm{C}_1,\cdots,\bm{C}_M\in \mathbb{R}^{d\times K}$ respectively, the method is exactly the additive quantization \cite{babenko2014additive}. Minimizing the quantization error $\sum_{\bm{x}\in X}||\bm{x}-\sum_{i=1}^M\bm{c}_{i}(\bm{x})||_2^2$ can obtain all the dictionaries and the selected columns w.r.t. each $\bm{x}$. Guo and Kumar et al. \cite{guo2016quantization} propose the product quantization (PQ) based method (QUIP). They partition each data point $\bm{x}\in \mathbb{R}^d$ into $M$ equal parts (i.e. $\bm{x}=[(\bm{x}^{(1)})^T; \cdots ; (\bm{x}^{(M)})^T]^T$ ) and quantize each part $\bm{x}^{(i)}$  and the corresponding codebook $\bm{C}_i\in \mathbb{R}^{\frac{d}{M}\times K}$ independently. The approximation of $\bm{x}$ is the concatenation of selected column (denoted as $\bm{c}_i(\bm{x})$) from each codebook, i.e. $\bm{x}\approx [\bm{c}_1(\bm{x})^T; \cdots ; \bm{c}_M(\bm{x})^T]^T$. Wu and Guo et al. \cite{wu2017multiscale} reveal that the data points with high variance norm degrade the performance of PQ based methods for MIPS problem. To address this problem, they propose the Multi-scale quantization method, i.e. "VQ-PQ" framework. The PQ is carried out on the normalized residual error of the vector quantization (VQ). Further more, Guo and Sun et al. \cite{guo2020accelerating} propose the the anisotropic quantization loss, i.e. the total loss is the weighted sum of parallel loss and orthogonal loss. Combining the "VQ-PQ" framework with the anisotropic quantization loss, they provide the state-of-the-art quantization based open source (ScaNN) for MIPS problem. The paper ~\cite{10.1145/3447548.3467441} even extends the quantization technique to online mode.

Greedy search on Delaunay Graph guarantees the optimal solutions for NNS problem. However, constructing exact Delaunay Graph is unacceptable for large scale datasets because of its time complexity. So, constructing approximate Delaunay Graph attracts the researchers' attentions. The Navigable Small World graph (NSW) \cite{malkov2014approximate} algorithm constructs the graph by inserting the node one by one and relies on the small world property to improve the relevance among vertices. 
The Hierarchical Navigable Small World graph (HNSW) \cite{malkov2018efficient} is the improved version of NSW. HNSW distributes nodes on different layers and nodes on the upper layer are the subset of nodes on the lower layer. HNSW algorithm presents excellent accuracy and efficiency for metric measure function in practice.
Inspired by the Delaunay Graph, some researches generalize it from metric space to non-metric space. The authors of paper \cite{morozov2018non} define the $s$-Delaunay Graph on non-metric space. They even prove that the typical greedy search on $s$-Delaunay graph guarantees the optimal solutions for queries. However, the construction of exact $s$-Delaunay graph also suffers the curse of  high dimensionality and the large scale of dataset. So they compromise to construct the  approximate $s$-Delaunay Graph. Especially, they set the inner product as the non-metric measure function and construct the approximate $s$-Delaunay Graph by referring the NSW method. The constructed graph is called as ip-NSW which is the generalized version of NSW. Their empirical studies show that the greedy search or the beam search on the ip-NSW performs very well for MIPS problem. Tan and Zhou et al. \cite{tan2019efficient} propose IPDG algorithm which adjusts the edge selection strategies of ip-NSW, by skipping the dispensable candidate neighbors when constructing the graph. Their experiments present that the sparse graph IPDG leads to high efficiencies for MIPS problem. For fast inner product search on graph, Zhou and Tan et al. \cite{zhou2019mobius} propose a new method to construct the proximity graph for MIPS problem by Mobius transformation. Concretely, the key step is to make Mobius transformation for each data point $\bm{x}$, i.e. $\bm{x}=\bm{x}/||\bm{x}||^2$. Then the transformed dataset is used to construct a graph. Especially, for database with large scale and high dimensionality, the paper proposes a method to construct graph with the help of an auxiliary zero vector. Their experiments show that their method has efficient performance at the same recall compared with other methods. These graph based methods have shown the strong performance for MIPS problem and NNS problem. In this paper, we also focus on graph based technique for MIPS problem. Especially, our main attention is to investigate how to learn to search on the graph. In fact, we construct proximity graph by referring the ip-NSW graph, IPDG graph and Mobius graph. Then, we propose a learnable model to route on these proximity graphs.

Learning to search is widely used for solving structured prediction problems like Part of Speech Tagging \cite{chang2015learning}, Machine Translation \cite{negrinho2018learning}, Scene Labelling \cite{cheng2017stacked}, recommendation systems ~\cite{liu2021reinforced,wang2022multi} and so on. Baranchun and Persiyanov et al. \cite{baranchuk2019learning} propose an imitation learning based method to learn to route on the proximity graph for MIPS problem. At training stage, the ground truths of all training queries are required, i.e. we need to find the real optimal solution of \textbf{E.q. (\ref{definition:mips})} for each training query before training the agent. The method forces an agent to walk toward the ground truth at each step such that the length of routing path is as short as possible and the probability to reach the ground truth is as larger as possible. One of the drawbacks of this method is that it requires the ground truths for all training queries. Usually, the larger the database is, the more training queries are needed. However, finding all the ground truths for large scale training queries is unavailable. Their method is inappropriate for large scale database. In this paper, we propose a reinforcement based method which can train the agent even if the ground truths are lacking so that our algorithm is more practical for large scale database. Besides, if there are a few demonstrations collected by experts (i.e. we get a small number of ground truths or a few shortest paths from initial vertices to the ground truths), our algorithm can also utilize these demonstrations to improve the agent's routing policy. Overall, our algorithm combines the reinforcement learning with imitation learning and utilizes the information as much as possible.

\section{Reinforcement Model}
As the shortage of labeled data, we use reinforcement learning to solve MIPS problem. Concretely, we want to train a smart agent to route on the proximity graph automatically by reinforcement learning. For MIPS problem, the agent can find a good solution on the graph for a given query.  Firstly, we introduce the components of our reinforcement model. Secondly, we introduce the structure of the agent used in our model.

\subsection{Components of the reinforcement model}
To better understand the reinforcement model, we suppose the proximity graph $G$ is known. The reinforcement model can be represented by the finite-state Markov decision process (MDP) $M=\{\mathcal{S},\mathcal{A},\mathcal{T},\gamma,\mathcal{R}\}$. $\mathcal{S}$ is the state space. In this paper, the agent's locations on the proximity graph make up the state space. Concretely, each vertex on the graph denotes a location (i.e. a state), such that there are $n$ states where $n$ is the number of vertices on the graph. For convenient usage, we don't discriminate the state $\bm{s}$ and its corresponding vertex's vector, i.e. we use the original vector of vertex to represent the state. $\mathcal{A}$ is the action space. At different locations, the agent senses different neighbors. At certain location, the current action set is the edges which connect to the unvisited neighbors of current location. The agent chooses one edge to walk along and reach the next location. $\mathcal{T}$ is the state transition model, i.e. $\mathcal{T}=\{prob(\bm{s},\bm{a},\bm{s'})\}$ which consists of the probability that the agent moves to state $\bm{s}'$ from state $\bm{s}$ once the agent conducts action $\bm{a}$. Actually, the transition probability is definite. Once the agent decides to route along certain edge, the state will shift to the corresponding neighbor definitely. In the MDP tuple, $\gamma\in [0,1]$ is the discount factor of the cumulative reward. $\mathcal{R}$ is the specific reward distribution which will be elaborated in later section.

\subsection{Architecture of the agent}
We want the agent can make proper decision when routing on the proximity graph, i.e. it can choose a proper vertex to expand from candidate set or neighbors. If the agent makes decision based on the local information like the adjacent nodes and the visited items, it may make myopic decisions and can get into local trap.
If the agent can obtain the some global information of the graph, it can make better decisions to avoid local optimal when routing on the graph. Luckily, Graph Convolutional Network (GCN) \cite{zhou2018graph,kipf2016semi} makes it possible to sense the global structure and it is widely used in pattern recognition and data mining. Supposing the proximity graph $G$ is known, we introduce two basic concepts: Graph Convolution Layer and Graph Convolution Block. The two components are the key parts of our used GCN.  Let's review the layer-wise propagation rule \cite{kipf2016semi} of GCN. It is
\begin{equation}\label{convolution_layer}
\begin{aligned}
H^{(l+1)}=\sigma(\widetilde{D}^{-\frac{1}{2}}\widetilde{A}\widetilde{D}^{-\frac{1}{2}}H^{(l)}W^{(l)})
\end{aligned}
\end{equation}
Here, $\widetilde{A}=A+I_{N}$ is the adjacency matrix of the graph $G$ with added self-connections. $\widetilde{D}$ is the diagonal matrix such that $\widetilde{D}_{ii}=\sum_{j}\widetilde{A}_{ij}$ and $W^{(l)}$ is the trainable weight matrix. $\sigma (\cdot)$ is the activation function. $H^{(l)}$ and $H^{(l+1)}$ are the input and the output of current layer respectively. The whole graph will be feeded into GCN, i.e. $H^{(0)}=X$ where $X$ is the vertex matrix and each row represents a vertex of the graph. We call the layer working like equation (\ref{convolution_layer}) as Graph Convolution Layer \cite{kipf2016semi}. A Graph Convolutional Block contains one Graph Convolution Layer \cite{kipf2016semi}, the ELU nonlinear activation function and the fully connected layer. The residual connection goes through it and the block ends with layer normalization.
\begin{figure}[htbp]
  \centering
  \includegraphics[width=4.0in]{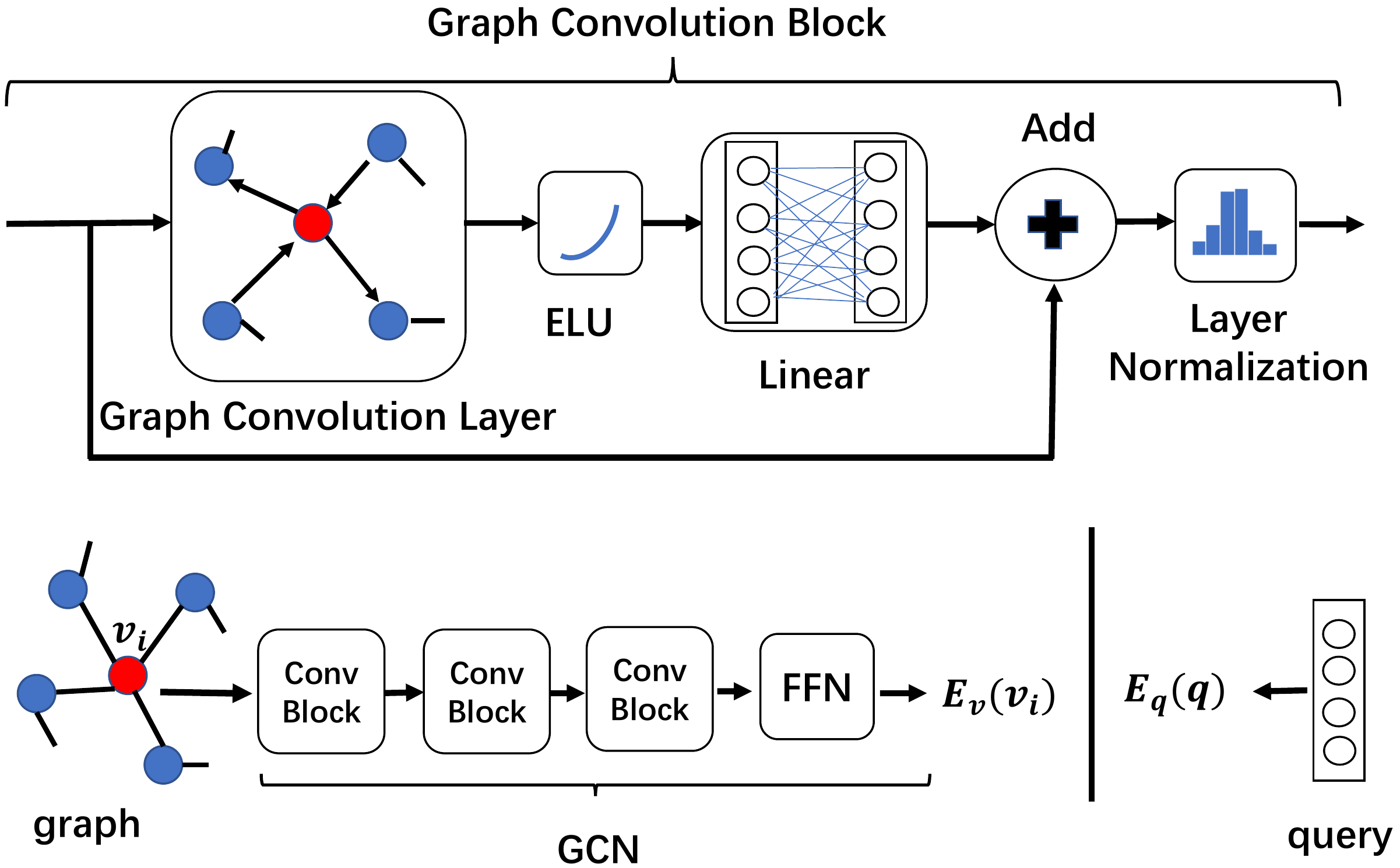}
  \caption{The architecture of the agent used in our experiments. Top part is the Graph Convolution Block which consists of a Graph Convolution Layer, the ELU nonlinearity and fully-connected layer. The residual connection goes through it.The bottom-left part is the GCN that is used to embed the vertices on graph into new space. It consists of three Graph Convolutions Blocks like the top part and followed by a feed-forward network (FNN) including two fully-connected layers with ELU nonlinearity. The bottom-right part is the embedding part for query.}
\label{fig:GCN}
\end{figure}
To embed the proximity graph, we use the same GCN architecture as that in paper \cite{baranchuk2019learning}. The GCN contains three Graph Convolution Blocks \cite{kipf2016semi} with ELU nonlinearity activation function as well as layer normalization \cite{ba2016layer} appended by residual connections \cite{he2016deep} for faster convergence. The original vertex matrix $X\in \mathbb{R}^{n\times d}$ will be embedded into a new matrix $X'\in \mathbb{R}^{n\times d'}$ by the GCN, where $n$ is the number of vertices, $d$ and $d'$ is original dimensionality and the new dimensionality respectively. Although query is not on the graph, we can also project the query into the same space with the embedded vertices, i.e. query need to be embedded to a new vector with dimensionality $d'$.

 We have illustrated the GCN framework in \textbf{Figure \ref{fig:GCN}}. The whole graph is the input of GCN (i.e. the bottom left part of \textbf{Figure \ref{fig:GCN}}) and the GCN will output the embedded matrix of the the original vertex matrix, i.e. each vertex on graph will be projected to a new vector. For convenient usage, $E_v(\cdot)$ denotes the embedded vector of vertex output by GCN and $E_q(\cdot)$ denotes the embedded vector of query, respectively. In our experiments, we project query $\bm{q}$ by identity transformation (i.e. $E_{q}(\bm{q})=\bm{q}$) if $d=d'$ and linear transformation (i.e. $E_q(\bm{q})=Wq$, $W\in \mathbb{R}^{d'\times d}$) if $d'\neq d$. It's easy to see that GCN will utilize the global information (i.e. $\widetilde{A}$ and $\widetilde{D}$ in \textbf{E.q. (\ref{convolution_layer})}) of the graph during the forward propagation. The global information of the graph is important to reduce the probability to get into local optimal trap. 
 
The GCN and the embedding part of query make up the agent. The agent can sense the global structure of proximity graph by GCN. The whole graph and a query are the input of the agent and the output of the agent is a vertex selected from a candidate set or neighbors, which will be elaborated in next section.

\section{Train the agent by reinforcement learning and imitation learning}
In this section, we elaborate the details about how to train the agent by reinforcement learning and imitation learning.
Firstly, we introduce the agent how to collect training instances when we give the agent some training queries. Secondly, we introduce the reward function which can utilize the labeled and unlabeled information. Then we introduce how to update the agent. Lastly, we introduce how to find the top $k$ items once the agent is trained.

\subsection{Collect training instances}
\begin{algorithm}[t]
   \caption{\textsc{Path collection} Algorithm}\label{alg:greedy}
   \textbf{Input:} The agent, the graph $G$, query $\bm{q}$, initial vertex $v_0$\\
   \textbf{Output:} a routing path\\
   \textbf{Process:}
\begin{algorithmic}[1]
   \STATE $V\leftarrow \{ v_0 \}$// the set of visited vertices\\
   \STATE $v\leftarrow v_0$
   \WHILE{has runtime budget}
       \STATE $Neighbors$ $\leftarrow$ the neighbors of $v$ on $G$
       \STATE $Candidates$ $\leftarrow$ $Neighbors \setminus V$
       \IF{$Candidates$ is not empty}
       \STATE $v\leftarrow$the agent select one vertex from $Candidates$
       \STATE $V\leftarrow  V\cup Candidates$
       \ELSE
       \STATE break
       \ENDIF
   \ENDWHILE
   \STATE \textbf{return} the routing path (i.e. the edges and nodes chosen by the agent in chronological order).
\end{algorithmic}
\end{algorithm}
In reinforcement learning, training instances are obtained by the interactions between the agent and the environment. We suppose the proximity graph and the agent are known. In order to train the agent, we collect the routing paths which the agent walks along on the proximity graph as the training instances once we give the agent some training queries. Next, we estimate the qualities of the collected paths, i.e. we need to verify whether the agent can find good solutions for training queries when it walks along these collected paths. Finally we train the agent to adjust its routing policy to get better paths. A natural way to collect routing paths is that the agent walks on the proximity graph again and again heuristically. In this paper, the routing framework which is used to collect training instances is similar to greedy strategy which is widely used at the classical search algorithms on graph. The details are elaborated in \textbf{Algorithm \ref{alg:greedy}}. The algorithm requires the agent, the proximity graph $G$, the query $\bm{q}$ and the initial vertex $v_0$ where the agent starts from. In the algorithm, the set $V$ (line 1) is the collection of the visited vertices which is initialized by $v_0$. At every step, the agent will choose one vertex to expand from the set of unvisited adjacent vertices of current vertex (line 4-9). When all the neighbors of current vertex have been visited or runtime budget is exhausted, the agent stops routing. The complete path is returned as one training instance (i.e. line 13). 

The key step for the agent is to make the decision that how to choose one node to expand from candidate set (i.e. line 7 in \textbf{Algorithm \ref{alg:greedy}}). We denote the vertices in candidate set as $\bm{c}_1,\dots,\bm{c}_m$ and the corresponding embedding vectors are $E_v(\bm{c}_1),\dots,E_v(\bm{c}_m)$ after embedding through GCN. The query is $\bm{q}$ and its embedding vector is $E_q(\bm{q})$. This selection strategy is based on the idea: if the original vectors have large inner products, then the corresponding embedding vectors also have large inner products and vice versa. To trade off exploration and exploitation, the agent samples one vertex from candidate set by softmax probability over the inner products between vertex embedding vectors and the query embedding vector. Concretely, each vertex $\bm{c}_i$ in the candidate set $Candidates$  is chosen to expand by the following probability
\begin{equation}\label{eq:softmax}
\begin{aligned}
P(\bm{c}_i|Candidates,\bm{q};\bm{\theta})=\frac{e^{\langle E_v(\bm{c}_i),E_q(\bm{q})\rangle/\tau}}{\sum_{\bm{c}\in Candidates}e^{\langle E_v(\bm{c}),E_q(\bm{q})\rangle/\tau}}
\end{aligned}
\end{equation}
where \bm{$\theta$} consists of the parameters of the agent and $\tau$ is the temperature to tune the smooth degree of the probability distribution. By the aforementioned content, the GCN helps the agent to sense the global information of the proximity graph. So the agent makes the decision based on global view of the graph by \textbf{E.q. (\ref{eq:softmax})}.

In order to improve the training efficiency, it should be avoided that feed the graph into the agent for each vertex embedding vector so that all the vertex embedding vectors should be pre-computed. The query embedding can be computed online because we only need to compute the query embedding one time during the routing process. 

\subsection{Assess the qualities of the training instances by reward function}
We can collect many training instances by \textbf{Algorithm \ref{alg:greedy}}. The qualities of the training instances reflect whether the agent makes a good decision or a bad decision at each step. In reinforcement learning, the reward function is used to quantify the qualities. For MIPS problem, our goal is to select one vertex on graph such that inner product with the query is maximized. So it is directed to set the reward function as
\begin{equation}\label{reward:no_gt}
\begin{aligned}
 r(\bm{s},\bm{a},\bm{s'})=\langle \bm{s'},\bm{q}\rangle-\langle \bm{s},\bm{q}\rangle.
 \end{aligned}
 \end{equation}
 Note that $\bm{q}$ is the query vector, $\bm{s}$ means the agent's current state, $\bm{s'}$ means agent's next state and action $\bm{a}$ means the agent chooses the edge that extends from $\bm{s}$ to $\bm{s'}$. Each state is represented by the corresponding vertex's vector by aforementioned content. So the reward function is the increment of inner product after the agent makes a decision. To do so, the naive cumulative reward of the routing path is $\langle \bm{s_t},\bm{q}\rangle-\langle \bm{s_0},\bm{q}\rangle$  where $\bm{s_t}$ is the terminal state of the path. This cumulative reward is the inner product of terminal state with query minus the initial one. It reflects the quality of the terminal state once the initial state $\bm{s_0}$ is fixed, i.e. it reflects the quality of the path. The reward function can push the agent to find the vertex whose inner product with query is as large as possible.

\begin{figure}[t!]\centering
\begin{minipage}[c]{0.20\linewidth}
\centering
    \includegraphics[width=1\linewidth]{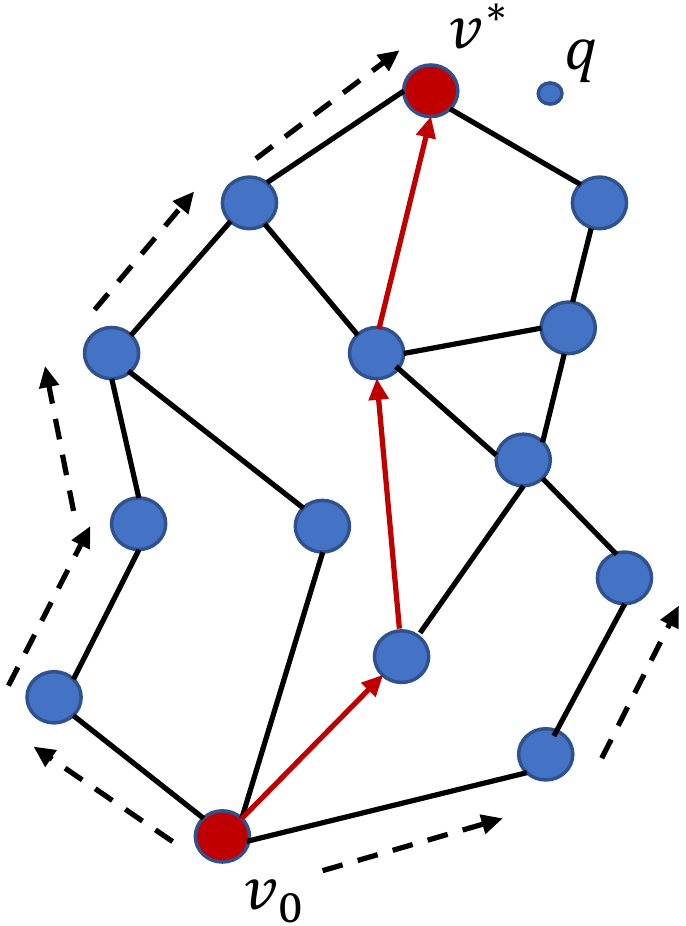}
\end{minipage}
\begin{minipage}[c]{0.20\linewidth}
\centering
    \includegraphics[width=1\linewidth]{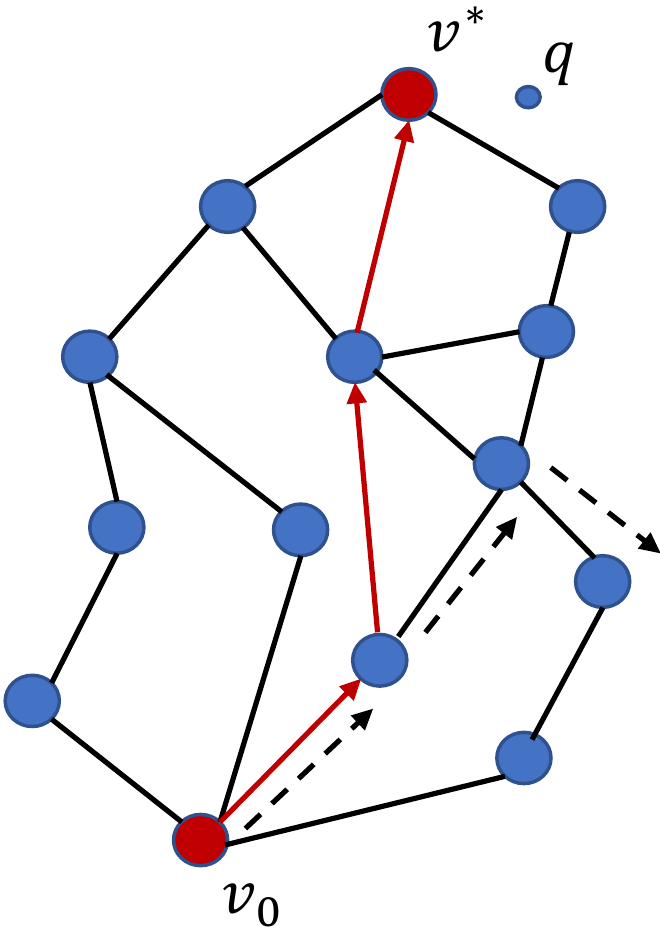}
\end{minipage}
\begin{minipage}[c]{0.20\linewidth}
\centering
    \includegraphics[width=1\linewidth]{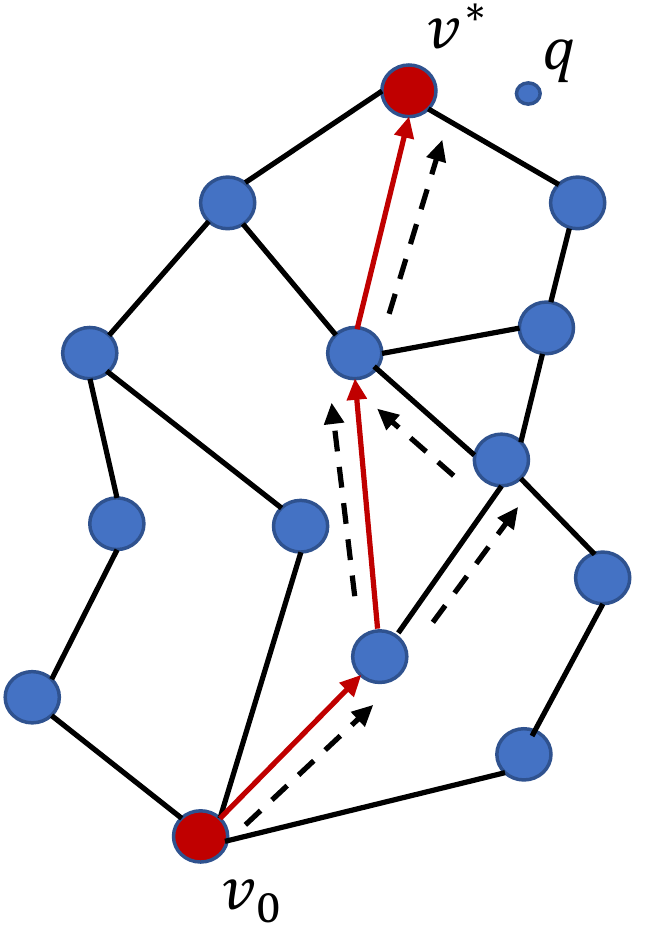}
\end{minipage}
\caption{$v_0,v^*$ are source node and target node, respectively. The flow of red arrows are shortest path and the flow of  black dash arrows are the routed path by the agent. \textbf{Left subfigure:} The agent routes along the path according to the naive reward function E.q.(\ref{reward:no_gt}). The collected path can be long (i.e. left dash arrows) or the agent can't reach the target node (i.e. right dash arrows); \textbf{Middle subfigure:} Only use the shaping function as reward. it causes the bad generalization ability. When the agent makes wrong decision at the second step, it can not correct the direction properly; \textbf{Right subfigure:} Use the combined reward function E.q. (\ref{reward:has_gt}). The agent can walk along the shortest path or it has the ability to correct it's decision at the third step when the agent makes wrong decision at the second step.}
\label{fig:path}
\end{figure}

The naive reward function \textbf{E.q. (\ref{reward:no_gt})} doesn't utilize any information of demonstrations and only considers the inner products. In such a naive way, the length of a routing path generated by the agent can be long or the agent can't reach the target node. The left subfigure of \textbf{Figure \ref{fig:path}} illustrates the drawbacks. We regard the node corresponding to the ground truth of the the training query as the agent's target node. The agent aims to route along the path whose length from source node to target node is as short as possible.
Calculating ground truth for each training query costs too much time. However, we can find the ground truths of a small number of training queries to guide the agent. Here, we will introduce the concept 'reward shaping' \cite{ng1999policy} to utilize these ground truths.

Suppose $v^*$ is the target node w.r.t. the ground truth of query $\bm{q}$ found by brute force search or other existed algorithms. $L(v_0,v^*)$ denotes the length of shortest path from source node $v_0$ to target node $v^*$, which can be obtained by Dijkstra algorithm or Breadth-First Search (BFS) on the graph. When $v^*$ is known for query $\bm{q}$, we want the agent can reach $v^*$ by following the shortest path. Then we adjust naive the reward function as
\begin{equation}\label{reward:has_gt}
\begin{aligned}
r(\bm{s},\bm{a},\bm{s'})&=\langle \bm{s'},\bm{q}\rangle -\langle \bm{s},\bm{q}\rangle-\alpha*(\gamma\cdot L(\bm{s'},v^*)-L(\bm{s},v^*))
\end{aligned}
\end{equation}
$\alpha \ge 0$ is the trade-off hyper-parameter. For undirected graph, $L(\bm{s'},v^*)-L(\bm{s},v^*)$ can only have three possible values, i.e. $\{-1,0,1\}$. If the agent follows the shortest path from state $\bm{s}$ to the target node $v^*$, then $L(\bm{s'},v^*)-L(\bm{s},v^*)=-1$.  If the agent walks away from the $v^*$, $L(\bm{s'},v^*)-L(\bm{s},v^*)=1$. Otherwise, $L(\bm{s'},v^*)-L(\bm{s},v^*)=0$, i.e. the agent 
takes a non-meaningful step and the length of the shortest path to the target node doesn't change. For directed graph, $L(\bm{s'},v^*)-L(\bm{s},v^*)$ is at least -1.  Intuitively, we will give the agent additional reward when it follows the shortest path to the target. Conversely, we will punish the agent when it walks away from the target. However, we hope that reward shaping of \textbf{E.q. \textbf{(\ref{reward:has_gt})}} doesn't affect the optimal policy of the original reward function \textbf{E.q. \textbf{(\ref{reward:no_gt})}} because our aim is to find the vertex that maximizes the inner product with query. Luckily, Andrew Y Ng and Harada et al. \cite{ng1999policy} have provided concrete theoretical basis. We need to adjust \textbf{E.q. \textbf{(\ref{reward:has_gt})}} mildly that set $L(\bm{s_t},v^*)=0$ where the $\bm{s_t}$ denotes the terminate vertex of the training path.

In our experiments, we will give the ground truths for a small number of training queries and the agent regards the corresponding nodes as targets. So the reward function has three cases: (a). the agent has no target; (b). the agent has target and the next state will be terminal state; (c). the agent has target and the next state is not the terminal state. Overall, we can declare the complete reward function as follows.
\begin{equation}\label{reward:overall}
r(\bm{s},\bm{a},\bm{s'})=
\begin{cases}
\langle \bm{s'},\bm{q}\rangle-\langle \bm{s},\bm{q}\rangle &\text{(a)} \\

\langle \bm{s'},\bm{q}\rangle -\langle \bm{s},\bm{q}\rangle+\alpha * L(\bm{s},v^*) &\text{(b)}\\

\langle \bm{s'},\bm{q}\rangle -\langle \bm{s},\bm{q}\rangle-\alpha *(\gamma L(\bm{s'},v^*)-L(\bm{s},v^*)) &\text{(c)}
\end{cases}
\end{equation}
For convenience, we just use $r(\bm{s},\bm{s'})$ to replace $r(\bm{s},\bm{a},\bm{s'})$ because there is only one edge between $\bm{s}$ and $\bm{s'}$ on our proximity graph. Next, we introduce the existed theoretical basis of our reinforcement model.
\begin{Theorem}[Theorem 1 of \cite{ng1999policy}]\label{Theo:maintheorem}
Let any $\mathcal{S},\mathcal{A},\gamma$ and any shaping reward function $F:\mathcal{S}\times \mathcal{A}\times \mathcal{S}\rightarrow \mathbb{R}$ be given. We say $F$ is a potential-based shaping function if there exists a real-value function $\Phi : \mathcal{S}\rightarrow \mathbb{R}$ such that for all $\bm{s}\in \mathcal{S}-\{\bm{s_T}\}, \bm{a}\in \mathcal{A},\bm{s'}\in \mathcal{S}$
\begin{equation}\label{eq:potentian_fucntion}
\begin{aligned}
F(\bm{s},\bm{a},\bm{s'})=\gamma \Phi (\bm{s'})-\Phi (\bm{s})
\end{aligned}
\end{equation}
(where $\mathcal{S}-\{\bm{s_T}\}=\mathcal{S}$ if $\gamma \le 1$ and $\bm{s_T}$ is the terminal state and $\Phi (\bm{s_T})=0$). Then $F$ is a potential-based function and is a necessary and sufficient condition for it to guarantee consistency with the optimal policy when learning from $M'=\{\mathcal{S},\mathcal{A},\mathcal{T},\gamma, \mathcal{R}+F\}$.
\end{Theorem}
Concretely speaking, when the shaping function satisfies the form of \textbf{E.q. (\ref{eq:potentian_fucntion})}, adding the shaping function to the reward function won't change the optimal policy of the MDP model. Obviously, our shaping function satisfies the form of \textbf{E.q. (\ref{eq:potentian_fucntion})}, i.e. $\Phi (\bm{s})=-\alpha L(\bm{s},v^*)$ when the reward is added with the shaping function. The shaping function can help the agent to collect high-quality paths as the training instances. If we only use the \textbf{E.q. (\ref{reward:no_gt})} as reward function, the collected paths can be too long or the agent can't reach the target node. If we only use the shaping function as the reward, i.e. force the agent walks along the shortest paths, this can lead to bad generalization ability. If the agent makes a wrong step, it's hard to correct the direction. Combining the naive reward \textbf{E.q. (\ref{reward:no_gt})} with the shaping function, the collected paths keep the the inner products of the nodes on the paths with queries as large as possible and the lengths of the paths as small as possible. \textbf{Figure \ref{fig:path}} illustrates the advantages of our reward function \textbf{E.q. (\ref{reward:has_gt})} intuitively.

\subsection{Update the policy of the agent}
The agent's policy should be updated according the the collected training instances and the specified reward function. In fact, it is to update the agent's parameter vector $\bm{\theta}$. We use the classical policy gradient method, \textsc{Reinforce} , to train our agent. For \textsc{Reinforce}, subtracting a baseline from the returned  reward leads to reduction in variance and allows faster convergence. In our paper, The \textsc{Reinforce} has self-critic baseline. Recall that the agent uses the following probability distribution to sample a candidate to expand when it collects the training instances by \textbf{Algorithm \ref{alg:greedy}}.
\begin{displaymath}
\begin{aligned}
P(\bm{c_i}|Candidates,\bm{q};\bm{\theta})=\frac{e^{\langle E_v(\bm{c}_i),E_q(\bm{q})\rangle/\tau}}{\sum_{\bm{c}\in Candidates}e^{\langle E_v(\bm{c}),E_q(\bm{q})\rangle/\tau}}
\end{aligned}
\end{displaymath}
Denote the sampled node as $\bm{c}$ and obtain reward $r(\bm{s},\bm{c})$, where $\bm{s}$ and $\bm{c}$ denote the current state and next state respectively.  we sample another $b$ nodes by the same distribution, denoted as $\bm{c_1},\bm{c_2},\cdots,\bm{c_b}$ and the corresponding rewards are $r(\bm{s},\bm{c_1}),\cdots,r(\bm{s},\bm{c_b})$. We set the average reward $\frac{1}{b}\sum_{i=1}^b r(\bm{s},\bm{c_i})$ as the baseline, i.e. $b(\bm{s})=\frac{1}{b}\sum_{i=1}^b r(\bm{s},\bm{c_i})$. Suppose that the length of current training path is $N$ and the nodes of the path are $\bm{s_1},\cdots, \bm{s_N}$ sequentially. The cumulative discounted reward is
\begin{equation}\label{eq:discunt_cumulative_reward}
\begin{aligned}
G_t=\sum_{i=t}^{N}\gamma ^{i-t}(r(\bm{s_{t-1}},\bm{s_t})-b(\bm{s_{t}})).
\end{aligned}
\end{equation}
Let $\bm{\theta}$ denote the parameter vector of the agent. $\eta$ denotes the learning rate which is a hyper-parameter. $\pi(\bm{a}|\bm{s};\bm{\theta})$ denotes the agent's policy which is probability that agent conducts action $\bm{a}$ at state $\bm{s}$. 
Then we can get  $\pi(\bm{a}|\bm{s};\bm{\theta})=P(\bm{c}|Candidates,\bm{q};\bm{\theta})$ where $Candidates$ is the unvisited neighbors of $\bm{s}$ (i.e. line 5 in \textbf{Algorithm \ref{alg:greedy}}) and $\bm{a}$ is the edge between $\bm{s}$ and $\bm{c}$. Finally, we update the policy by gradient ascent as follows
\begin{equation}
\begin{aligned}
\bm \theta \leftarrow \bm \theta +\eta G_t\bigtriangledown \pi(\bm{a_t}|\bm{s_t},\bm{\theta}).
\end{aligned}
\end{equation}
\subsection{Top $k$ search on the graph}
\begin{algorithm}[t]
   \caption{\textsc{Beam Search} Algorithm}\label{alg:beamsearch}
   \textbf{Input:} Graph $G$, query $\bm{q}$, initial vertex $v_0$, integer $k$\\
   \textbf{Output:} A set consisting of $k$ vertices\\
   \textbf{Process:}
\begin{algorithmic}[1]
   \STATE $V\leftarrow \{ v_0 \}$// the set of visited vertices\\
   \STATE $Candidates\leftarrow \{v_0\}$
   \WHILE{has runtime budget \& $Candidates\neq \Phi$ }
       \STATE $c\leftarrow$ select one vertex from $Candidates$
       \STATE $Neighbors$ $\leftarrow$ the neighbors of $c$ on $G$
       \FOR{$v\in Neighbors$}
       \IF{$v\notin V$}
       \STATE $V\leftarrow V\cup \{v\}$
       \STATE $Candidates\leftarrow Candidates\cup \{v\}$
       \ENDIF
       \ENDFOR
   \ENDWHILE
   \STATE \textbf{return} Top $k$ vertices of $V$
\end{algorithmic}
\end{algorithm}

If the training process is completed, we need to test the performance of the trained agent. The classical greedy search is easy to be stuck in local optimum. To test the quality of the agent, we utilize a more roust framework, beam search (\textbf{Algorithm \ref{alg:beamsearch}}), which is a typical way to navigate on proximity graph. The algorithm requires the graph $G$, query $\bm{q}$, the initial vertex $v_0$ and size of solution set $k$. We use $V$ (line 1) to collect the visited vertices during the searching process and $Candidates$ (line 2) consists of candidate vertices which can be chosen to expand. At very step, the agent chooses a vertex from $Candidates$ (line 4) and senses its adjacent vertices (line 5). For every adjacent vertex of the chosen candidate, it will be added into set $V$ and $Candidates$ if it hasn't been visited (line 6-10). When the runtime budget is exhausted or there is no vertex can be chosen to expand in $Candidates$, the search process stops. At last, the top $k$ items of visited set $V$ will be returned (line 13). In fact, the embedding vector $E_v(v)$ for each node $v$ of the graph can be pre-computed once the agent is trained. In line 4 of \textbf{Algorithm \ref{alg:beamsearch}}, we select a candidate which maximizes the inner product with query embedding vector from candidate set, i.e. 
\begin{equation}
    \bm{c}=\arg\max_{\bm{c'}\in Candidates} \langle E_q(\bm{q}),E_v(\bm{c'})\rangle.
\end{equation}
The beam search can be a general search framework which is not only used for our agent. If there is no agent, we can also search top $k$ items on the proximity graph. Concretely, we can select a vertex from candidate set by $\bm{c}=\arg\max_{\bm{c'}\in Candidates} \langle \bm{q},\bm{c'}\rangle$ (line 4 of beam search). 

Either to collect training instances (\textbf{Algorithm \ref{alg:greedy}}) or to find the top $k$ items (\textbf{Algorithm \ref{alg:beamsearch}}), we has the parameter runtime budget. In practice, the absolute runtime depends on the the quality of codes and the devices such as CPU and GPU and so on. In our experiments, we ignore other cost and regard the computation times of inner product as the runtime budget.
\begin{algorithm}[htb]
   \caption{\textsc{ip-NSW Construction } Algorithm \cite{morozov2018non}}\label{alg:nsw}
   \textbf{Input:} Database $X$, similarity function $\bm{s(a,b)}=\langle \bm{a},\bm{b}\rangle$, maximum vertex degree $M$\\
  \textbf{Output:} Graph $G$\\
   \textbf{Process:}
\begin{algorithmic}[1]
   \STATE Initialize graph $G=\Phi$\\
   \FOR{$\bm{x}$ in $X$ }
       \STATE $S$=\{$M$ nodes from $G$ , s.t. the corresponding vectors $\bm{y}$ give the largest values of $s(x,y)$\}
       \STATE Add $\bm{x}$ to the graph $G$ and connect it by the bi-directional edges with vertices in $S$.
       \FOR{Each node of S}
            \IF{The degree of the node is larger than M}
             \STATE Only retain the top-M neighbors whose inner products with the node is maximum.
             \ENDIF
       \ENDFOR
   \ENDFOR
   \STATE \textbf{return} Graph $G$
\end{algorithmic}
\end{algorithm}
\section{Proximity graph construction}
In the former sections, we suppose the proximity graph is known. In the section, we introduce the proximity graphs used in our experiments.
The ideal proximity graph is the s-Delaunay graph but the large time complexity to construct the s-Delaunay graph makes the researchers compromise to construct approximate s-Delaunay graph. The experiments are conducted on three proximity graphs, i.e. ip-NSW graph ~\cite{morozov2018non}, IPDG graph ~\cite{tan2019efficient} and Mobius graph ~\cite{zhou2019mobius}. We will elaborate the details of these graphs in this section.

\subsection{ip-NSW graph}
The classical NSW \cite{malkov2014approximate} and the Hierarchical NSW (HNSW) \cite{malkov2018efficient} are designed for NNS problem. ip-NSW ~\cite{morozov2018non} is the extension for the non-metric measure function, i.e. inner product. We describe the construction process of ip-NSW graph by \textbf{Algorithm \ref{alg:nsw}}. The input consists of the database matrix $X$, the similarity function $s(\bm{a}, \bm{b})=\langle \bm{a},\bm{b}\rangle$ and the maximum degree of the node on the constructed graph. Firstly, the graph is empty (line 1) and each data point of $X$ is inserted into the graph iteratively. For the data point $\bm{x}$, the M vertices (denoted as $S$) which have maximum inner products with $\bm{x}$ are collected from current graph $G$ (line 3). This step can be accomplished by Beam Search (i.e. \textbf{Algorithm \ref{alg:beamsearch}}). Then, the data point $\bm{x}$ is inserted into the graph and the edges are added between $\bm{x}$ and the nodes in $S$ (line 4). If the degree of node in $S$ exceeds the limitation $M$, only the top-M neighbors can be retained and the redundant edges will be eliminated (line 7). After all the data points are inserted into the graph, the algorithm outputs the final graph (line 11).

\subsection{IPDG graph}
\begin{algorithm}[htb]
   \caption{\textsc{IPDG Construction } Algorithm \cite{tan2019efficient}}\label{alg:ipdg}
   \textbf{Input:} Database $X$, similarity function $\bm{s(a,b)}=\langle \bm{a},\bm{b}\rangle$, candidate size $N$, maximum vertex degree $M$\\
   \textbf{Process:}
\begin{algorithmic}[1]
   \STATE Initialize the graph $G=\Phi$, round=0\\
   \WHILE{round<2}
        \STATE round=round+1
        \FOR{$\bm{x}$ in $X$ }
           \STATE $v_0\leftarrow$ choose a vertex from $G$ randomly
           \STATE $C\leftarrow$ \textsc{Greedy Search}($G$,$\bm{x}$,$v_0$,$N$,$N$,$s(\cdot,\cdot)$)\ \  //{i.e. find $N$ vertices for $\bm{x}$ w.r.t. $\bm{s(x,\cdot)}$ on current graph}
           \STATE $B\leftarrow$ Neighbor Selection($C,M$)\ \  //{i.e. select $M$ vertices from the $N$ vertices}
           \STATE Add edges $\vec{\bm{xy}}$ to $G$ for every $\bm{y}\in B$
           \FOR{$\bm{y}\in B$}
                \STATE $C\leftarrow \{\bm{z}\in X: \vec{\bm{yz}}\in G\} \cup \{\bm{x}\}$\ \ //{i.e. get the neighbors of $\bm{y}$ and $\bm{x}$ itself}
                \STATE $D\leftarrow$\textsc{Neighbor Selection}($C,M$)\ \ //{i.e. select $M$ vertices from $C$}
                \STATE Remove the out-going edge of $\bm{y}$ and add $\vec{\bm{yz}}$ to $G$  for every $\bm{z}\in D$
           \ENDFOR
        \ENDFOR
    \ENDWHILE
   \STATE  \textbf{return} Graph $G$
\end{algorithmic}

 \textbf{Sub-procedure:} \textsc{Neighbor Selection}(C,M)
\begin{algorithmic}[1]
\STATE $B=\Phi$
\FOR{$\bm{y}\in C$}
    \IF{$\bm{y}^T\bm{y}\ge \max_{\bm{z}\in B}\bm{y}^T\bm{z}$}
        \STATE $B\leftarrow B\cup \{\bm{y}\}$
    \ENDIF
    \IF{$|B|\ge M$}
        \STATE $B\leftarrow B\cup \{\bm{y}\}$
    \ENDIF
\ENDFOR
\STATE \textbf{Return:} B
\end{algorithmic}
\end{algorithm}
The construction procedure of IPDG is a 2-round algorithm which is illustrated  in \textbf{Algorithm \ref{alg:ipdg}}. The first round can construct a graph and the second round can refine the graph. The input of \textbf{Algorithm \ref{alg:ipdg}} consists of the database matrix $X$, the similarity function $s(\bm{a},\bm{b})=\langle \bm{a},\bm{b}\rangle$, an integer $N$ which limits the size of candidates and the maximum degree $M$. In the first round, the initial graph is empty and each data point of $X$ is inserted into the graph iteratively. For each data point $\bm{x}\in X$, we need to find $N$ nodes from current graph by greedy search as the candidate neighbors of $\bm{x}$ (line 6). The greedy search is elaborated by \textbf{Algorithm \ref{alg:greedysearch}} which finds $N$ nodes by breadth-first-search on the graph. Only $M$ of the $N$ candidate nodes can by selected as the neighbors of $\bm{x}$ (line 7). Selecting neighbors from candidate set is elaborated by the sub-procedure function, i.e. Neighbor Selection function (Sub-procedure of \textbf{Algorithm \ref{alg:ipdg}}). Then, $\bm{x}$ is inserted into the graph and the directed edges from $\bm{x}$ to its neighbors (line 8) are added into the graph. Finally, the neighbors of the selected candidates are updated (line 9-13). Repeat the procedure one more time at the second round to refine the graph. Lastly, the algorithm outputs the graph (line 16).

\begin{algorithm}[t]
   \caption{\textsc{Greedy Search} Algorithm}\label{alg:greedysearch}
   \textbf{Input:} Graph $G=(V,E)$, query $\bm{q}$, initial vertex $v_0$, size of candidate set $N$, integer $k$, similarity function $s(\cdot,\cdot)$\\
\begin{algorithmic}[1]
   \STATE Mark $v_0$ as checked and the rest vertices as unchecked
   \STATE $C\leftarrow \{v_0\}$ // the candidates set
   \WHILE{some vertices unchecked}
       \STATE $C\leftarrow C\cup{ \{v\in V: u\in C , v\ unchecked, (u,v)\in E\}}$
       \STATE Mark the vertices in $C$ as checked
       \STATE $C\leftarrow$Top $N$ vertices of $C$  in descending order of $s(\bm{q},\bm{x})$ for $\bm{x\in C}$
       \IF{$C$ converges}
       \STATE break
       \ENDIF
   \ENDWHILE
   \STATE \textbf{return} Top $k$ vertices of $C$ in descending order of $s(\bm{q},\bm{x})$ for $\bm{x\in C}$
\end{algorithmic}
\end{algorithm}
\subsection{Mobius graph}
The ip-NSW graph and IPDG graph are constructed based on the original database. However, Mobius graph is constructed based on the Mobius transformed data with the help of an auxiliary zero vector. The used Mobius transformation is $\bm{y}=\bm{x}/||\bm{x}||^2$ for $\bm{x}\in X$. We illustrate the construction procedure in \textbf{Algorithm \ref{alg:mobius}}.  Mobius transformation is applied to each data point $\bm{x}\in X$ (line 1). The transformed data points and a zero vector make up the new database (denoted as $Y$). The construction is conducted on the new database. The initial graph is a full connected graph with the first $M$ vertices of $Y$ (line 3). The left points of $Y$ will be inserted into the graph iteratively. For each $\bm{y}_i\in Y$ ($i\ge M$), the greedy search (i.e. \textbf{Algorithm \ref{alg:greedysearch}}) is applied to find $N$ candidate neighbors for $\bm{y}_i$ (line 5). Note that the similarity function is negative $l_2$-distance, i.e. $s(\bm{a},\bm{b})=-||\bm{a}-\bm{b}||$ in line 5. $M$ neighbors are chosen from the $N$ candidates by the sub-procedure (i.e. Neighbor Selection function, line 6). Then, $\bm{y}_i$ is inserted into the graph and connected to the selected neighbors (line 7). For each neighbor of $\bm{y}_i$, its neighbor also need to be updated (line 8-12). After all the data points of $Y$ are inserted into the graph, we need to remove $\bm{y}_0$ and the relevant edges from the graph because there is no corresponding data point in the original database for $\bm{y}_0$. The original data points $\bm{x}_i$ ($1\le i\le n$) are used to replace the $\bm{y}_i$ on the graph (line 14). Lastly, the algorithm outputs the final graph (line 15).
\begin{algorithm}[htb]
   \caption{\textsc{Mobius Construction } Algorithm \cite{tan2019efficient}}\label{alg:mobius}
   \textbf{Input:} Database $X$, candidate size $N$, maximum vertex degree $M$\\
   \textbf{Output:} Graph $G$\\
\begin{algorithmic}[1]
    \STATE $n\leftarrow |X|$. For $i\in \{1,2,\dots,n\}$, $\bm{y}_i=\bm{x}_i/||\bm{x}_i||^2$;
    \STATE $Y=\{\bm{y}_0,\bm{y}_1,\dots,\bm{y}_n\}$, $\bm{y}_0=\bm{0}$;
   \STATE $G\leftarrow $ full connected graph with vertices   $\{\bm{y}_0,\bm{y}_1,\dots,\bm{y}_{M-1}\}$;
    \FOR{$i=M$ to $n$ }
       \STATE $C\leftarrow$ \textsc{Greedy Search}($G$,$\bm{y}_i$,$\bm{y}_0$,$N$,$N$, negative $l_2$-distance)
       \STATE $B\leftarrow$ Neighbor Selection($\bm{y}_i,C,M$)\ \  //{i.e. select $M$ vertices from the $N$ vertices}
       \STATE Add edges $(\bm{y}_i,\bm{z})$ to $G$ for every $\bm{Z}\in B$
       \FOR{$\bm{z}\in B$}
            \STATE $C\leftarrow \{\bm{w}\in Y: (\bm{z},\bm{w})$ is an edge of G$\} \cup \{\bm{y}_i\}$
            \STATE $D\leftarrow$\textsc{Neighbor Selection}($\bm{z},C,M$)\ \
            \STATE Let nodes of $D$ be the out-neighbors of $\bm{z}$ in G
       \ENDFOR
    \ENDFOR
    \STATE Remove $\bm{y}_0$ and its incident edges from $G$ and replace the node of $G$ by the ones before transformation.
   \STATE  \textbf{return} Graph $G$
\end{algorithmic}

 \textbf{Sub-procedure:} \textsc{Neighbor Selection}($\bm{x}$,C,M)
\begin{algorithmic}[1]
\STATE $B=\Phi$
\STATE Order $\bm{y}_i\in C$ in ascending order of $||x-\bm{y}_i||$
\STATE $i\leftarrow1$
\WHILE{$|B|\le M$ and $i\le |C|$}
    \IF{$||\bm{x}-\bm{y}_i||\le \min _{\bm{z}\in B}||\bm{z}-\bm{y}_i||$}
        \STATE $B\leftarrow B\cup \{\bm{y}_i\}$
    \ENDIF
    \STATE $i\leftarrow i+1$
\ENDWHILE
\STATE \textbf{Return:} B
\end{algorithmic}
\end{algorithm}

\section{Experiments}
\begin{table*}[htbp]
\caption{Dataset}
\begin{tabular*}{11cm}{l|lllll}
\hline
Dataset &\#Base Data  &\#Ratings &\#Training &\#Validation &\#Test\\
\hline
MovieLens & 59,047& 25,000,095  & 130,032 &16,254 &16,255\\
Amazon &79,851 & 1,828,551 & 111,177 &13,907 &13,898\\
Echonest  & 60,654 & 4,422,471 &174,372&21,796 &21,798 \\
\hline
\end{tabular*}
\label{tab:dataset}
\end{table*}
In this section, we present the experimental evaluations compared with the latest and start-of-the-art algorithms for MIPS problem. All experiments are conducted on three common datasets: MovieLens 25M\footnote{http://grouplens.org/datasets/movielens/}( \textbf{MovieLens}), Amazon Movies and TV\footnote{http://jmcauley.ucsd.edu/data/amazon/} (\textbf{Amazon}) and \textbf{Echonest}\footnote{http://millionsongdataset.com/tasteprofile/}. For \textbf{MoviesLens}, it has 162,541 users, 59,047 items and 25,000,095 rating records. For \textbf{Amazon},
 users that have rated at least 5 items and items at least rated by 5 users can be retained respectively. 138,972 users, 79,851 items and 1,828,551 rating records are left. For \textbf{Echonest}, users that have rated at least 5 items can be left and 217,966 users, 69,654 items, 4,422,471 ratings are retained. For each dataset, we use the alternating least square method w.r.t. implicit feedback dataset of \cite{hu2008collaborative, Takcs2011ApplicationsOT} (The open source code is available.\footnote{https://github.com/benfred/implicit}) to factorize the item-user rating matrix as user embedding vectors and item embedding vectors with 96 dimensions. The item embedding  vectors are regarded as base data points and the user embedding vectors are regard as queries. For the user embedding vectors, we randomly split each of them into three parts which 80\% of users are regarded as the training queries, 10\% of users are used as validation queries and the left 10\% of users are test queries. For our algorithm, we normalize each dataset in the way that each item vector is divided by the mean $l_2$ norm of the all the item vectors and each query is normalize as unit vector. The pre-process about the data points and queries will not affect the solution of MIPS problem. The summaries of datasets are presented in \textbf{Table \ref{tab:dataset}}.

 As the primary performance measure, we use Recall M@N, which is calculated as
\begin{equation}
\begin{aligned}
Recall\ M@N =\frac{\sum_{\bm{q}}|R(\bm{q})\cap T(\bm{q})|}{\sum_{\bm{q}}|T(\bm{q})|}
\end{aligned}
\end{equation}
where $R(\bm{q})$ is the top $N$ candidate items w.r.t. query $\bm{q}$ returned by algorithms and $T(\bm{q})$ is the true top $M$ items w.r.t. query $\bm{q}$. We compare the recall for all algorithms within the same running time, higher recall means better algorithm. Our algorithms have runtime budget at both training stage and test stage. We set the maximal number of inner product computations (IPC) as the budget. In fact, the programming language (e.g. C and C++ are more efficient than Python.), the programmers' skills (e.g. whether programmers use parallel computation mode) and the running devices will affect the efficiency of codes. For fair comparison, we ignore the absolute time but limit the IPC for all compared algorithms in later comparisons. All the hyper-parameters are tuned on validation queries. We demonstrate the used hyper-parameters firstly. $\gamma=0.9, \alpha=0.7,\tau=0.15,b=4$ for \textbf{MovieLens}; $\gamma=0.9, \alpha=1.0,\tau=0.5,b=4$ for \textbf{Amazon}; $\gamma=0.9, \alpha=2.7,\tau=1.0,b=4$ for \textbf{Echonest}. In the later subsection, we will demonstrate the hyper-parameter sensitivity experiments. Adam ~\cite{DBLP:journals/corr/KingmaB14} is used as the optimizer and learning rate is set to be $0.001$ with exponential decay. The comparison methods refer to the recommended parameters.  The maximum out degree for each vertex of all the proximity graphs is set to be 16 and more details about constructing the proximity graph can be found by \cite{morozov2018non,tan2019efficient,zhou2019mobius} for ip-NSW, IPDG and Mobius respectively. For all experiments, the batch size is 30 (i.e. 30 queries are used to train the agent at each time) and 60000 batches are used to train the agent iteratively.
\begin{table}[t]
\caption{Results of varying the dimension of query vector}
\begin{tabular}{|l|l|l|l|l|l|}
\hline
                             & \#dim64 & \#dim80 & \#dim96        & \#dim112 & \#dim128 \\ \hline
\multicolumn{1}{|c|}{MovieLens} & 0.668   & 0.533   & \textbf{0.757} & 0.368    & 0.293    \\ \hline
Amazon                       & 0.794   & 0.768   & \textbf{0.850} & 0.624    & 0.532    \\ \hline
Echonest                     & 0.884   & 0.854   & \textbf{0.951} & 0.741    & 0.656    \\ \hline
\end{tabular}
\label{table:embedding_query}
\end{table}
\subsection{Is it necessary to embed the queries?}
For each vertex on the proximity graph, it can be embedded into a new vector by the GCN of the agent. All the embedding vectors of the vertices can be obtained off line once the agent is trained. Similarly, we can also embed the queries into the same space with the embedding vertex vectors. However, embedding the queries must be online because we don't know the queries before we test the agent. In this subsection, we check whether we need to embed the queries. We embed the queries with dimensionality 96 into the new space with dimensionality 64, 80, 112, 128 respectively by the learnable linear transformation $E_q(\bm{q})=\textbf{W}\bm{q}$ ($\textbf{W}\in \mathbb{R}^{d'\times d}$) where $d$ is the original dimensionality and $d'$ is the new dimensionality. To train the agent, 30\% ground truths are provided for training queries. The ip-NSW is used as the proximity graph and the IPC equals to 256 if we don't embed the queries when collecting the training instances and conducting beam search on the graph. For fair comparison, we need to keep same computation budget whether we embed the queries or not. Embedding the queries by learnable linear transformation leads to extra computations so that IPC=$\frac{256\times d-d'\times d}{d'}$ ($d'\times d$ is the extra computations) if we embed the queries into new space. \textsc{Beam Search} (\textbf{Algorithm \ref{alg:beamsearch}}) is called to find the top $1$ and the results of Recall 1@1 are presented in \textbf{Table \ref{table:embedding_query}}.
\begin{figure}[htbp]
\centering
\begin{minipage}[c]{0.4\linewidth}
    \includegraphics[width=1\linewidth]{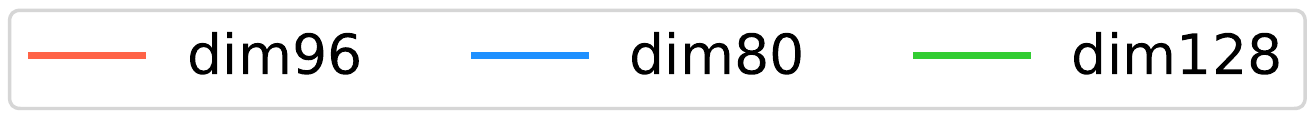}
\end{minipage}
\begin{minipage}[c]{0.40\linewidth}
    \includegraphics[width=1\linewidth]{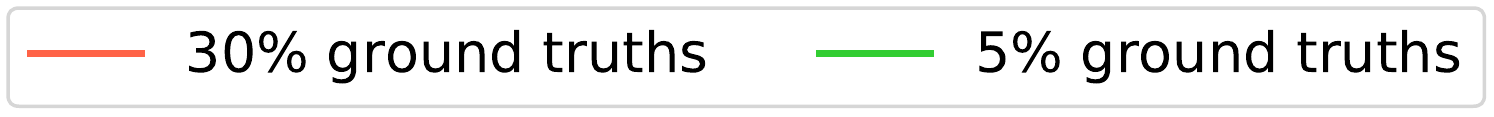}
\end{minipage}
\ \ \ \ \ \ \
\begin{minipage}[c]{0.40\linewidth}
    \includegraphics[width=1\linewidth]{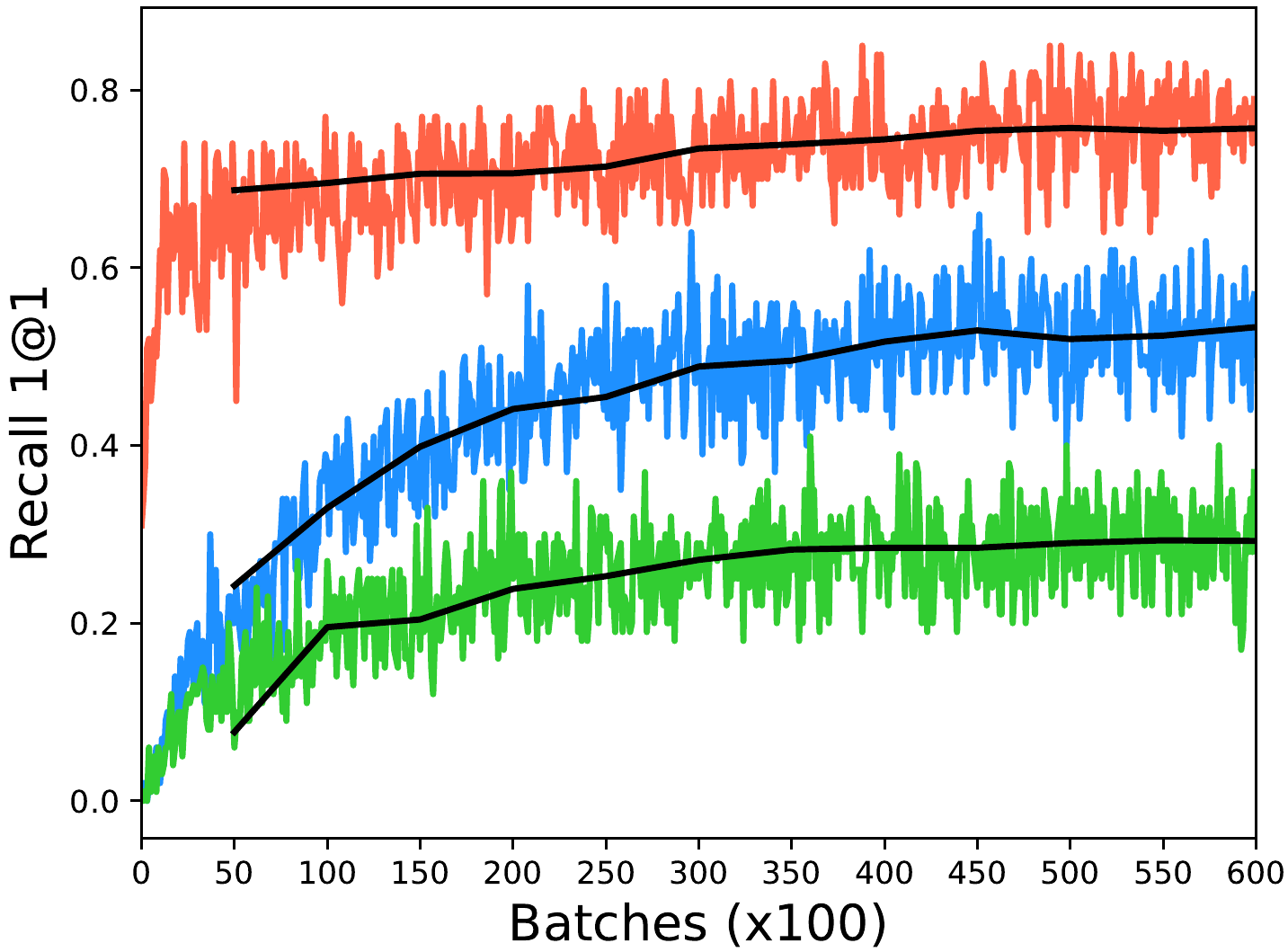}
\end{minipage}
\begin{minipage}[c]{0.40\linewidth}
    \includegraphics[width=1\linewidth]{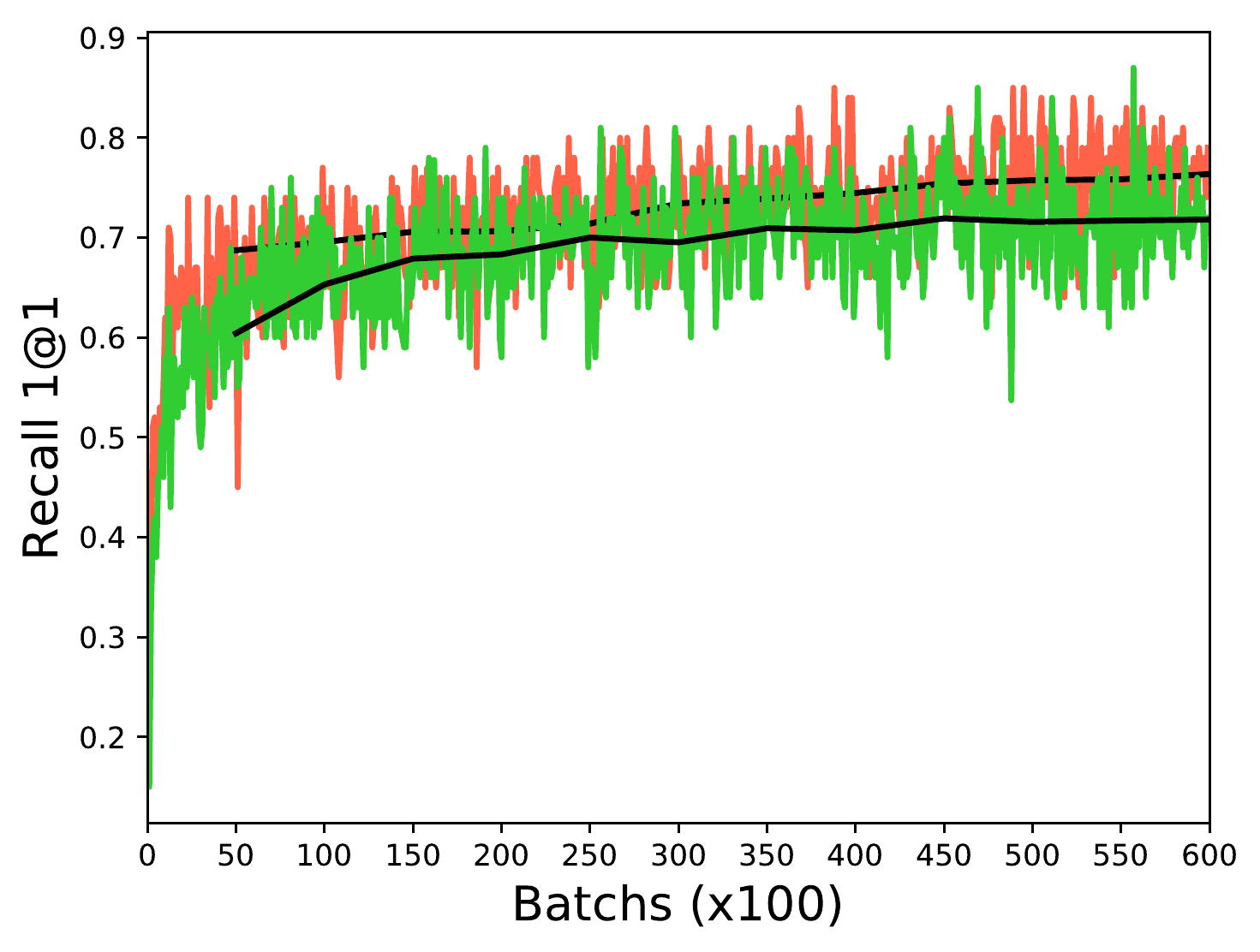}
\end{minipage}
\caption{Left: The convergence when varying the dimensionality. Right: The convergence when varying the ratio of ground truths. }\label{fig:convergence}
\end{figure}
From \textbf{Table \ref{table:embedding_query}}, we can know that it performs best if we don't embed the queries (i.e. the column \#dim96). It may be because embedding the queries will pull-in new parameter matrix $\textbf{W}$ but embedding queries by online mode can't make the agent to learn a good parameter matrix $\textbf{W}$. When the queries are embedded, the smaller $d'$ leads to higher recall. Especially on the dataset \textbf{Movies}, the degradation of recall is significant. This degradation of recall may be caused by two factors: the larger $d'$ leads to smaller IPC which means the agent explores less nodes on the graph; larger $d'$ also leads to more parameters in the learnable parameter matrix $\textbf{W}\in \mathbb{R}^{d'\times d}$ which is harder for the agent to learn it well. In fact, we've done more tests about the embedding of queries to verify our hypothesis, e.g. using a three-layer neural network to embed the queries 
can lead to the worse recall.  It's not necessary to embed the query by our results. In the left subfigure of \textbf{Figure \ref{fig:convergence}}, we present the convergence of recall with the increasing of training batches under different embedding dimensionality when the ratio of ground truths is 30\% on \textbf{MovieLens}. The zigzag lines w.r.t. red, blue and light green  are the estimated recall under 100 validation queries and the corresponding dark lines are the exact recall under all validation queries. The estimated recall is reported every 100 batches and the exact recall is reported every 5000 batches. We can know that our algorithm has good convergence whether we embed the query or not. The superiority of dim=96 (i.e. Don't embed the query) 
appears from initial time to the end. Embedding the query into lower dimensionality also exhibits this kind of superiority compared with embedding the query to higher dimensionality. In the following experiments, all the queries are embedded by identity transformation, i.e. $E_q(\bm{q})=\bm{q}$.

\subsection{Effectiveness of reward shaping}
\begin{figure}[t]
\centering
\subfigure{
\begin{minipage}[c]{0.32\linewidth}
\centering
    \includegraphics[width=1\linewidth]{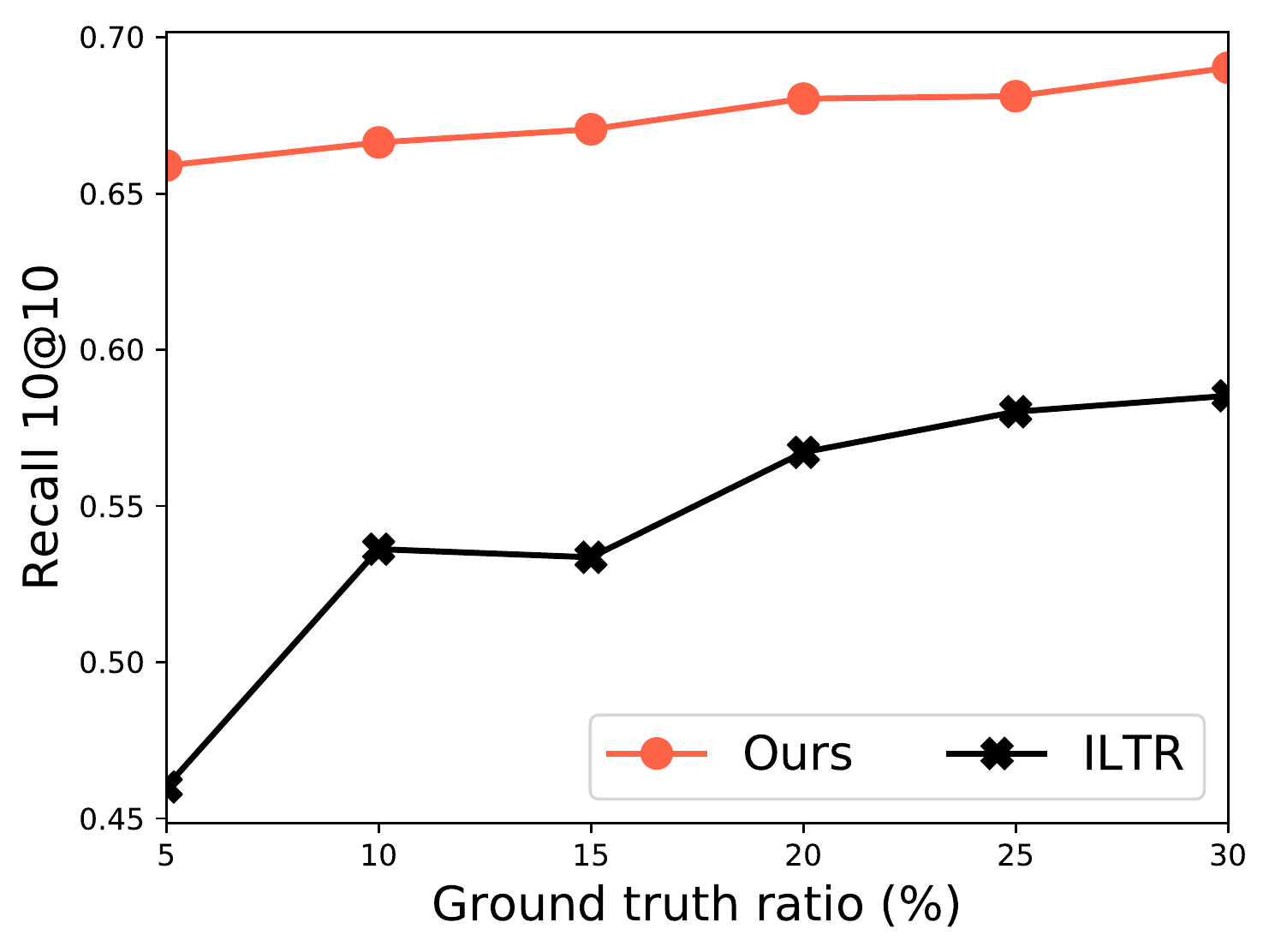}
    (a).MovieLens
\end{minipage}}
\subfigure{
\begin{minipage}[c]{0.32\linewidth}
\centering
    \includegraphics[width=1\linewidth]{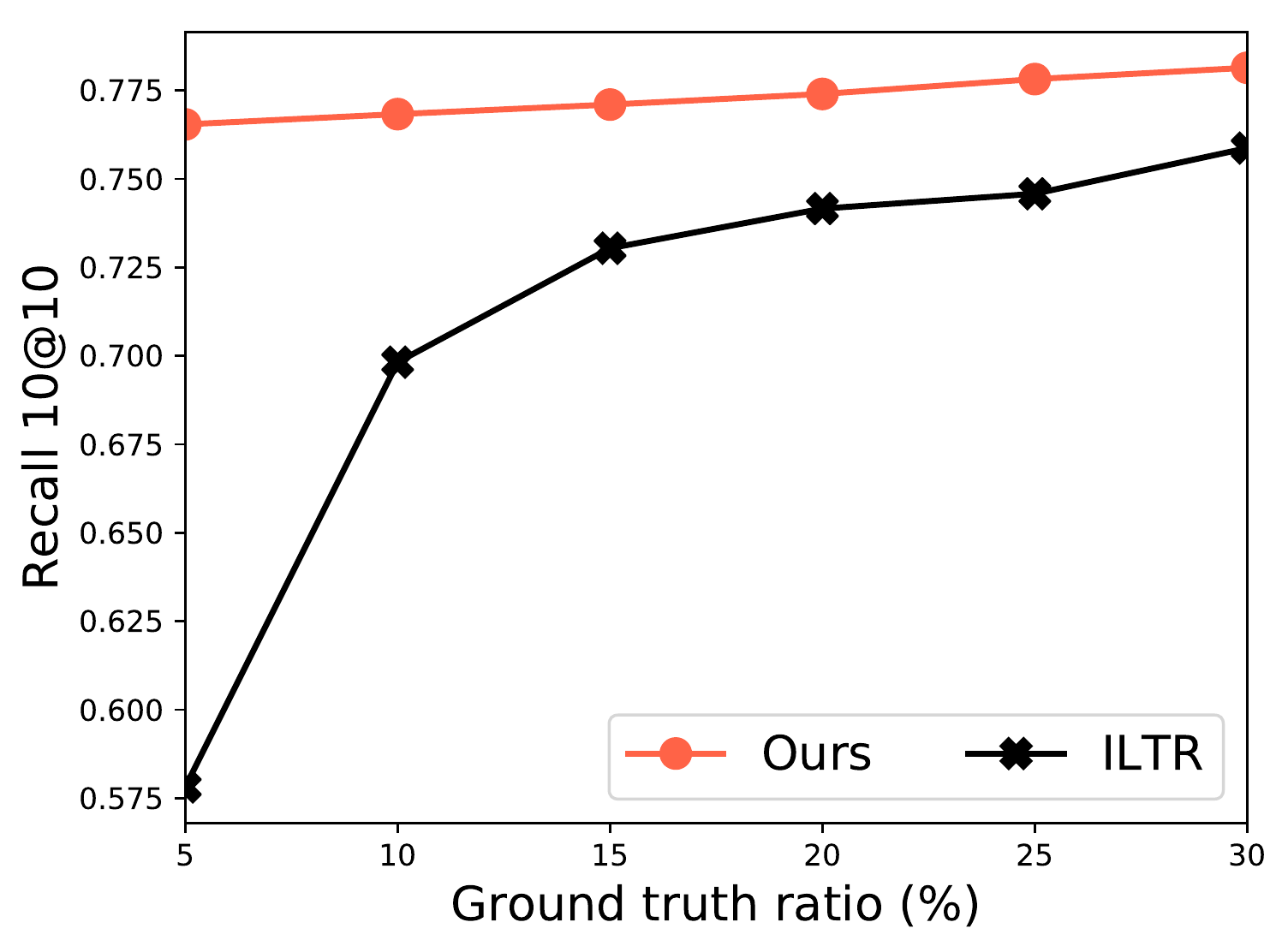}
    (b).Amazon
\end{minipage}}
\subfigure{
\begin{minipage}[c]{0.32\linewidth}
\centering
    \includegraphics[width=1\linewidth]{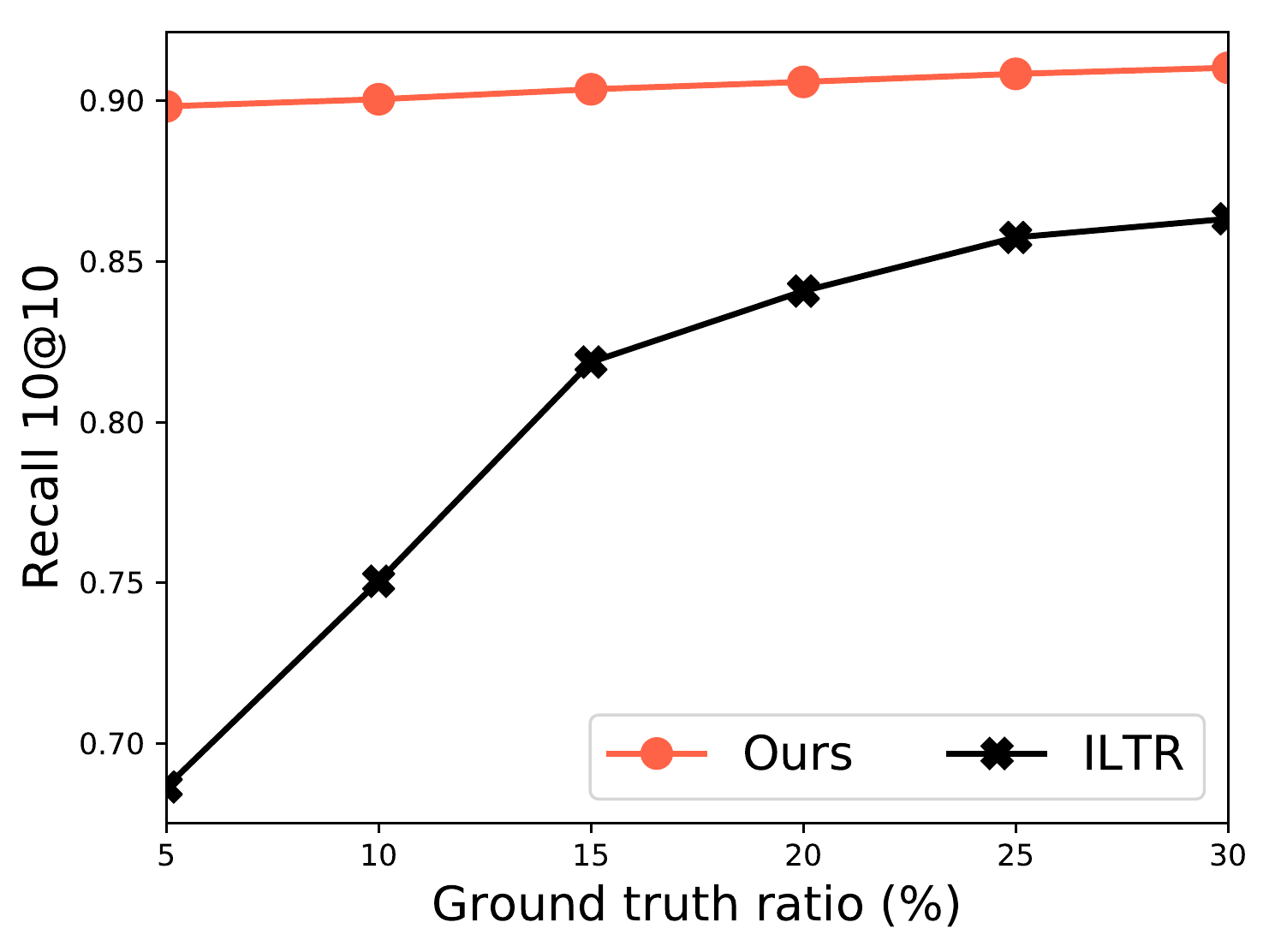}
    (c).Echonest
\end{minipage}}
\caption{The experimental results of varying the ratio of ground truths when we train the agent. }\label{fig:vary_gt_ratio}
\end{figure}

In this subsection, we verify the effectiveness of reward shaping. The only existed learnable algorithm to route on proximity graph for MIPS is proposed in \cite{baranchuk2019learning}. To train the agent, it requires the ground truths for all training queries. The agent imitates to follow the shortest path from source node to the target node at training stage. We abbreviate the algorithm as \textbf{ILTR}\footnote{https://github.com/dbaranchuk/learning-to-route}, i.e. imitation learning to route. Here, we construct the ip-NSW as the proximity graph for each dataset. For all training queries, we give 5\%, 10\%, 15\%, 20\%, 25\%, 30\% ground truths to train the agent, respectively. However, \textbf{ILTR} algorithm can only utilize the training queries with ground truths but our algorithm can utilize all the training queries. Going after the \textbf{ILTR}, Our algorithm also uses \textsc{Beam Search} (\textbf{Algorithm \ref{alg:beamsearch}}) to test the trained agent. IPC=256 when collecting the training instances and conducting beam search on the graph. $Recall\ 10@10$ is used to measure the performance. The experimental results are shown in \textbf{Figure \ref{fig:vary_gt_ratio}}.

Our algorithm always performs better than \textbf{ILTR} algorithm. Obviously, the advantage is more significant when the ground truths are inadequate. Especially, our algorithm even can be more than 50\% advantage when we just have 5\% ground truths on \textbf{MovieLens} dataset. As the recall 10@10 is large for our algorithm on \textbf{Echonest} dataset, the results seem smooth but still improve as the ratio of ground truths increase. So we can know that our algorithm is a good choice when we just have a small number of ground truths. As the increasing of ground truths, the performance of our algorithm improves constantly which means that some demonstrations for our agent can help it to make better decisions. We also present the convergence when varying the ratio of ground truths on \textbf{MovieLens} in the right subfigure of \textbf{Figure \ref{fig:convergence}}. Our algorithm still can converge even if there are only a few demonstrations. The advantage is continuous as the training batches increase if there are more ground truths.

\subsection{Compare with graph based methods}
\begin{table}[t]
\caption{The results on three kinds of graphs w.r.t. each dataset.}
\begin{tabular}{l|c|ll|ll|ll}
\multirow{2}{*}{Dataset} &
  \multicolumn{1}{l|}{\multirow{2}{*}{IPC}} &
  \multicolumn{2}{c|}{ip-NSW Graph} &
  \multicolumn{2}{c|}{IPDG Graph} &
  \multicolumn{2}{c}{Mobius Graph} \\ \cline{3-8} 
                           & \multicolumn{1}{l|}{} & Ours            & ip-NSW & Ours            & IPDG   & Ours            & Mobius \\ \hline
\multirow{3}{*}{MovieLens} & 128                   & \textbf{0.3502} & 0.2864 & \textbf{0.1886} & 0.0800 & \textbf{0.1383} & 0.0771 \\
                           & 256                   & \textbf{0.6943} & 0.6462 & \textbf{0.5939} & 0.3542 & \textbf{0.6541} & 0.1824 \\
                           & 512                   & \textbf{0.8962} & 0.8889 & \textbf{0.8115} & 0.7171 & \textbf{0.8836} & 0.6856 \\ \hline
\multirow{3}{*}{Amazon}    & 128                   & \textbf{0.5209} & 0.4901 & \textbf{0.3624} & 0.3062 & \textbf{0.2403} & 0.0523 \\
                           & 256                   & \textbf{0.7818} & 0.7596 & \textbf{0.6005} & 0.5495 & \textbf{0.5274} & 0.3295 \\
                           & 512                   & \textbf{0.9171} & 0.9074 & \textbf{0.7888} & 0.7573 & \textbf{0.6716} & 0.6188 \\ \hline
\multirow{3}{*}{Echonest}  & 128                   & \textbf{0.6897} & 0.6516 & \textbf{0.4303} & 0.3296 & \textbf{0.6121} & 0.4196 \\
                           & 256                   & \textbf{0.9120} & 0.8956 & \textbf{0.6346} & 0.5431 & \textbf{0.7541} & 0.6709 \\
                           & 512                   & \textbf{0.9847} & 0.9822 & \textbf{0.7902} & 0.7301 & \textbf{0.8106} & 0.7844
\end{tabular}
\label{table:other_graph_based_method}
\end{table}

\begin{figure}[t]
\centering
\subfigure{
\begin{minipage}[c]{0.32\linewidth}
\centering
    \includegraphics[width=1\linewidth]{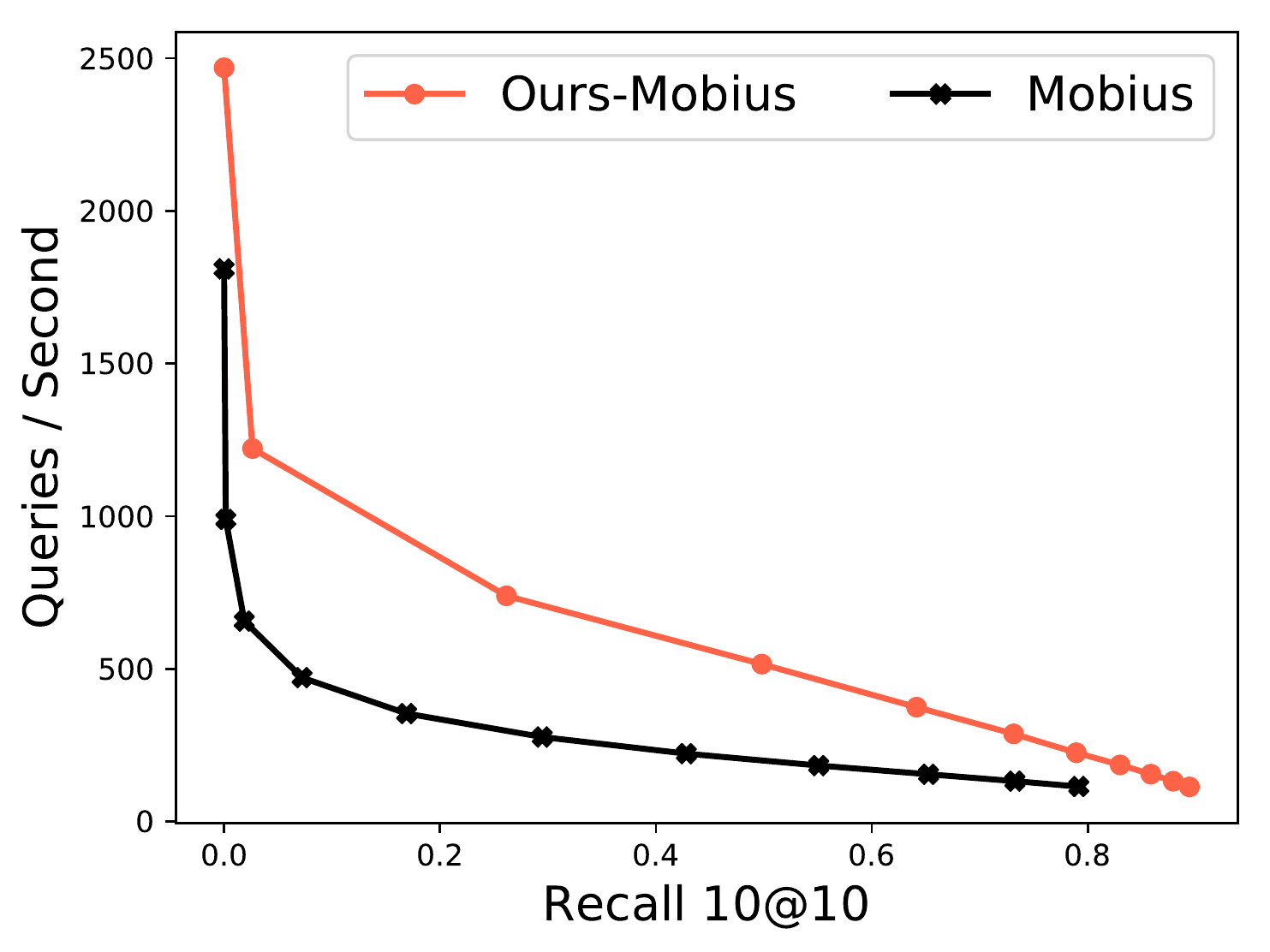}
    (a).MovieLens
\end{minipage}}
\subfigure{
\begin{minipage}[c]{0.32\linewidth}
\centering
    \includegraphics[width=1\linewidth]{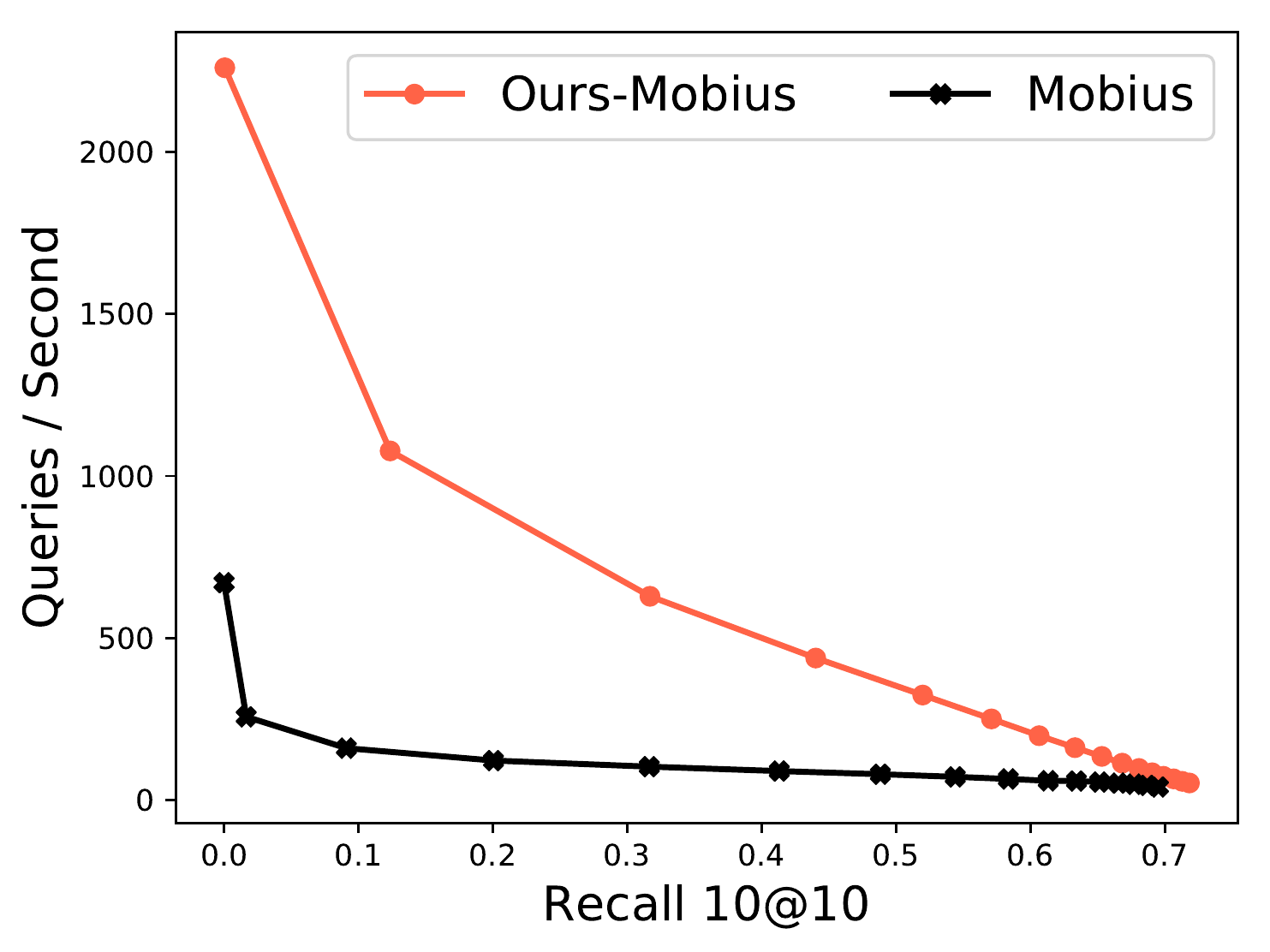}
    (b).Amazon
\end{minipage}}
\subfigure{
\begin{minipage}[c]{0.32\linewidth}
\centering
    \includegraphics[width=1\linewidth]{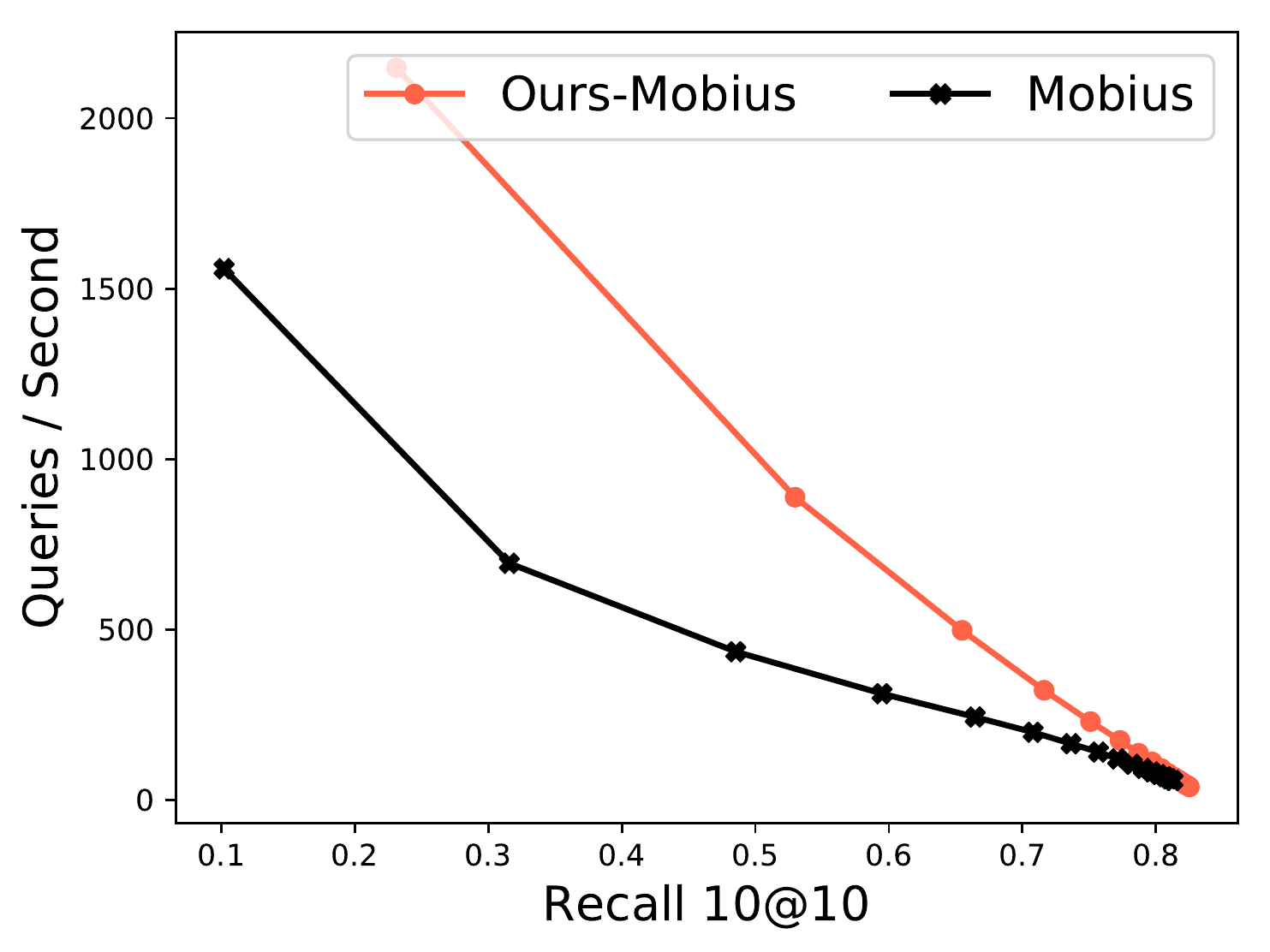}
    (c).Echonest
\end{minipage}}
\caption{The results of time vs recall on Mobius graphs of each dataset. }\label{fig:time_vs_recall}
\end{figure}
In this subsection, we compare our algorithm with ip-NSW algorithm ~\cite{morozov2018non}, IPDG algorithm~\cite{tan2019efficient} and Mobius algorithm ~\cite{zhou2019mobius}. Firstly, we need to construct the three kinds of graphs by the aforementioned content for each dataset. We call the graphs as ip-NSW graph, IPDG graph and Mobius Graph respectively. ip-NSW algorithm means that the beam search (i.e. \textbf{Algorithm \ref{alg:beamsearch}}) is conducted on the ip-NSW graph. 
The same goes for IPDG algorithm and Mobius algorithm. For our proposed algorithm, we train the agent on each graph and compare our algorithm with the corresponding algorithm on the same graph. The difference between our algorithm and the compared algorithm is that compared algorithm conducts beam search based on the the original vectors of nodes but our algorithm conducts beam search on the embedded vectors of nodes by the GCN part of the agent. It means that our algorithm conducts beam search on $E_v(v)$ where $v$ is the node of the graph. The IPC is set to be 256 when the agent collect training instances and 30\% ground truths of training queries are provided to train the agent. The IPC is set to be 128,256 and 512 when beam search is conducted on each graph respectively. Recall 10@10 is reported and he results are presented in \textbf{Table \ref{table:other_graph_based_method}}. We can know that our algorithm always performs best on each graph of each dataset. Especially, the improvements on IPDG graph and Mobius graph of each dataset are significant. These results indicate that the proposed  reinforcement model and the imitation learning model are effective to improve the performance of graph based methods for MIPS problem.  For MIPS problem, the efficiency is also important. We investigate the efficiency on Mobius graph for each dataset. The results are presented in \textbf{Figure \ref{fig:time_vs_recall}}. The vertical axis represents the number of tackled queries at each second and the horizontal axis represents recall 10@10. We can know that our algorithm is more efficient on each dataset. These results indicate that GCN in our proposed model really helps the agent to choose more proper candidates when beam search is conducted.

\subsection{Parameter sensitivity}
\begin{figure}[htbp]
\centering
\subfigure{
\begin{minipage}[c]{0.32\linewidth}
\centering
    \includegraphics[width=1\linewidth]{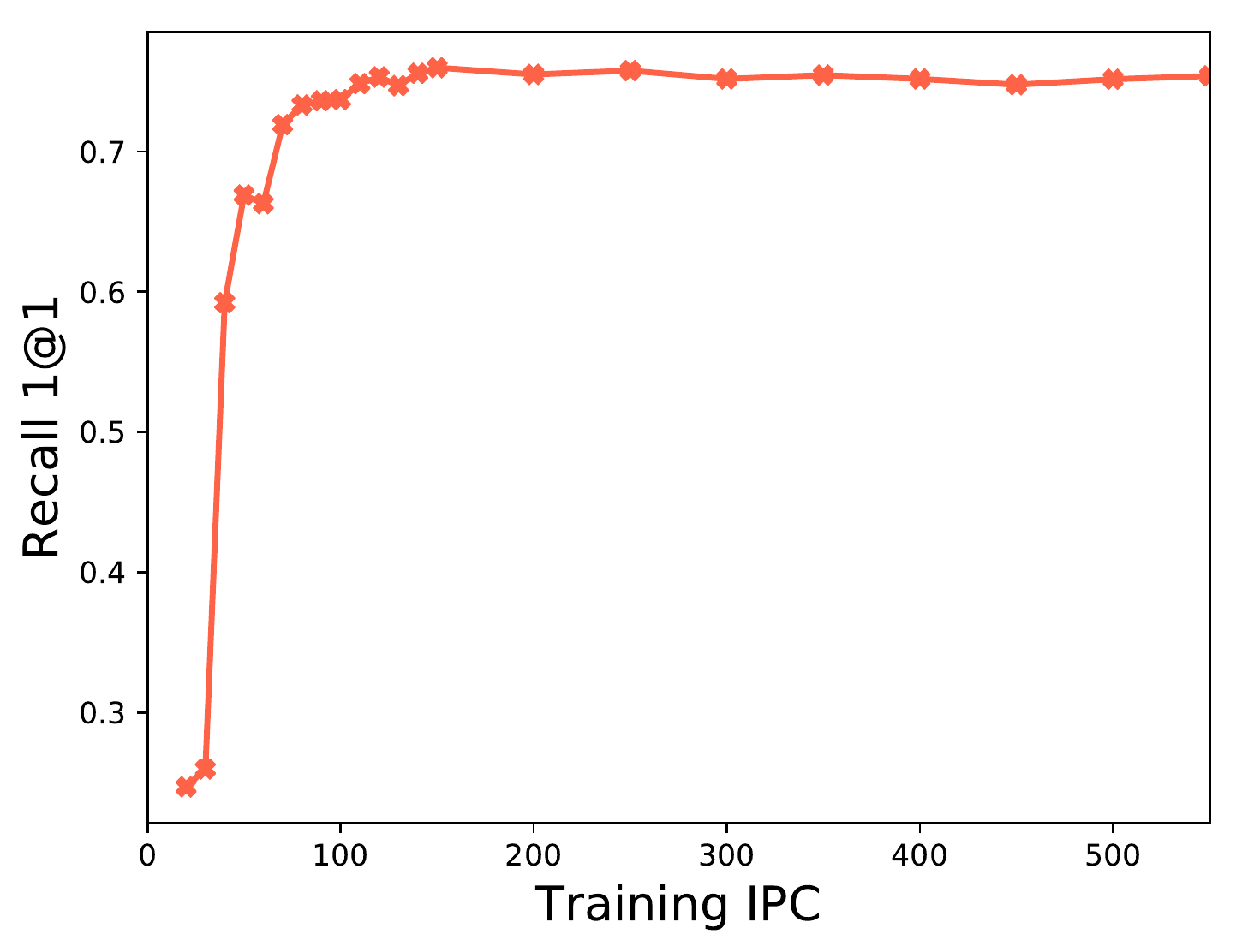}
    (a).MovieLens

\end{minipage}}
\subfigure{
\begin{minipage}[c]{0.32\linewidth}
\centering
    \includegraphics[width=1\linewidth]{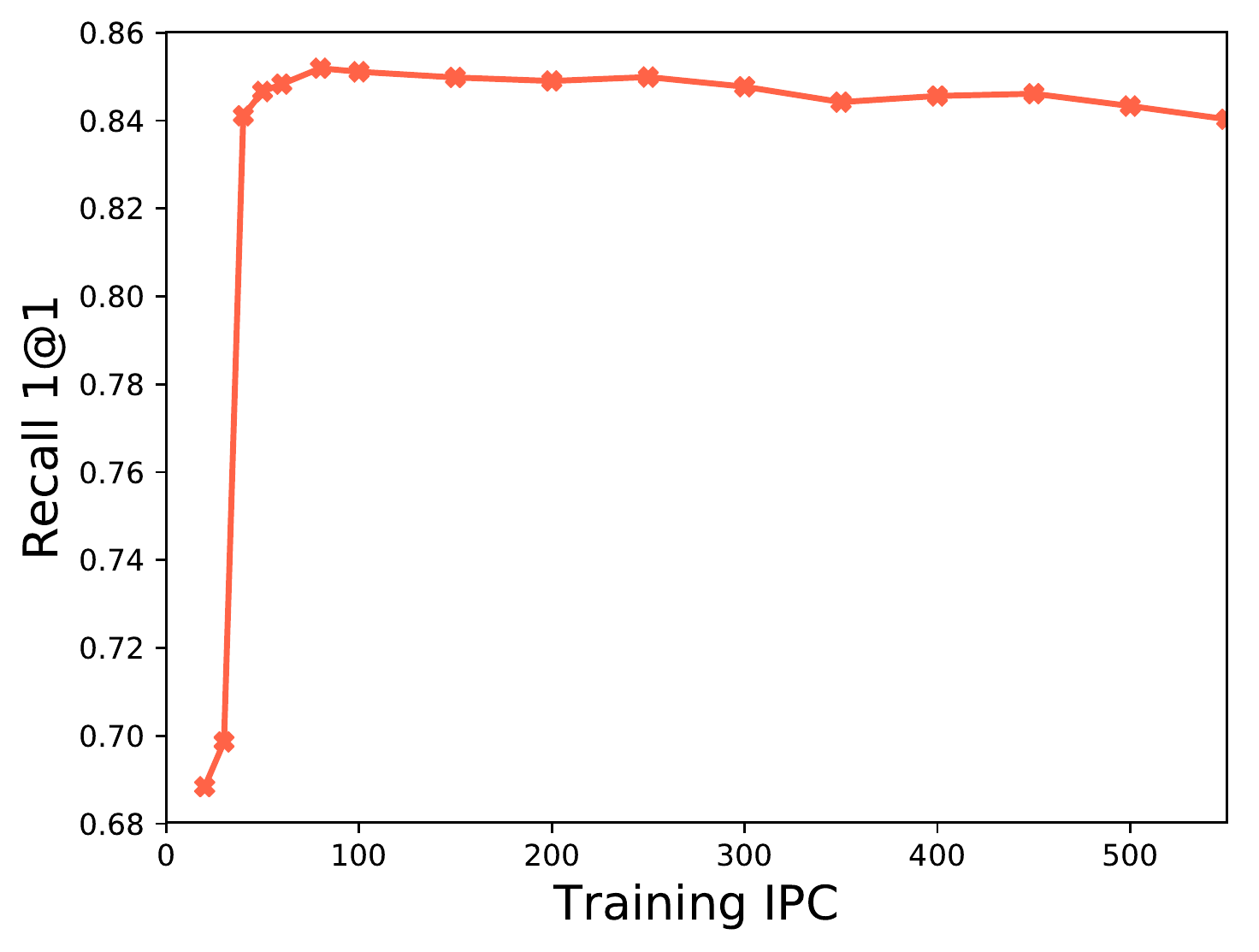}
    (b).Amazon
\end{minipage}}
\subfigure{
\begin{minipage}[c]{0.32\linewidth}
\centering
    \includegraphics[width=1\linewidth]{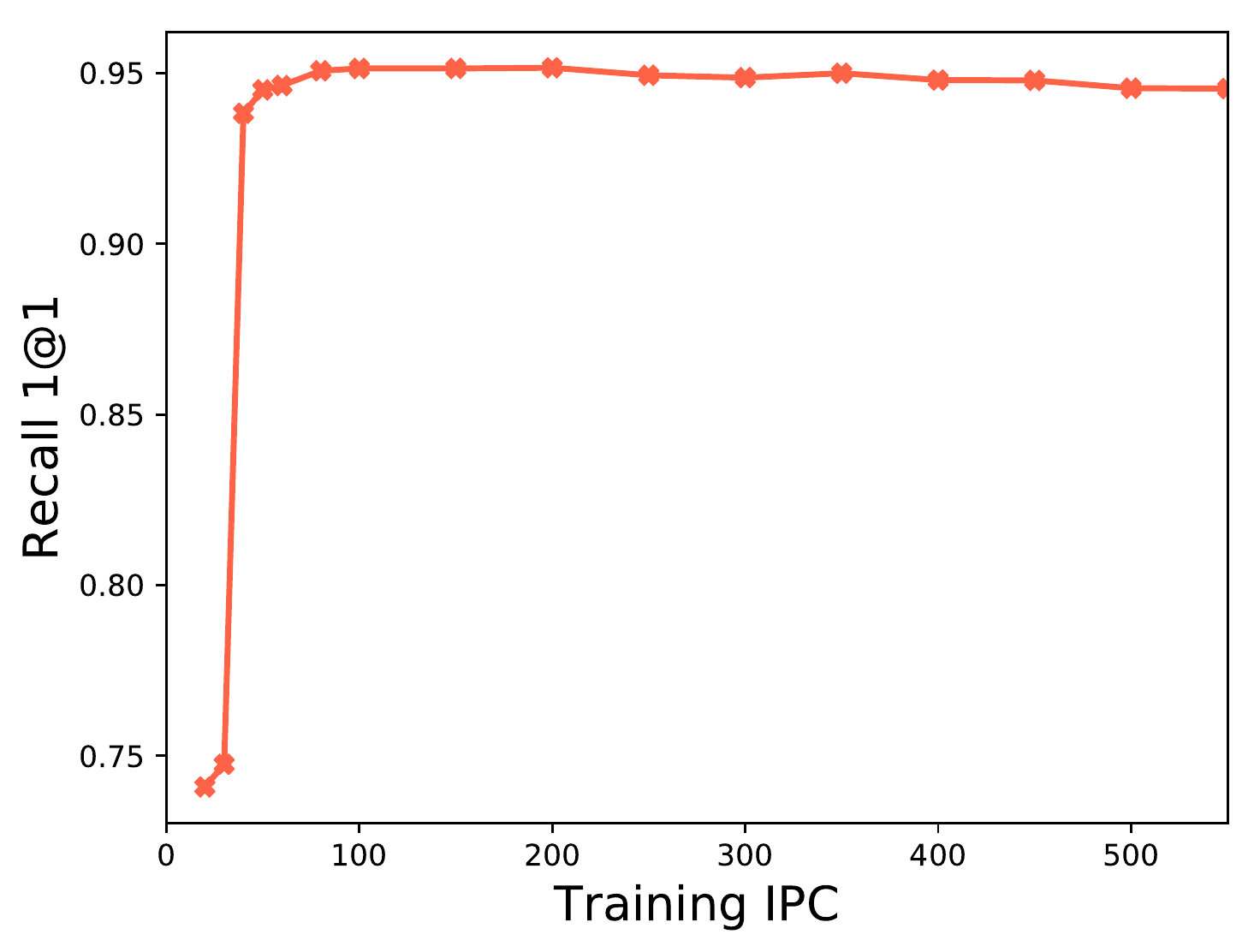}
    (c).Echonest
\end{minipage}}

\caption{The influence of runtime budgets when the agent collecting the training instances.}\label{fig:vary_training_dcs}
\end{figure}

When the agent collects training instances by \textbf{Algorithm \ref{alg:greedy}}, it needs to stop when the time of routing reaches the budget. Firstly, we see the influence of the runtime budget of collecting the training instances. The experiments are conducted on the ip-NSW graph of each dataset. The IPC is set to be 256 when we use beam search (i.e. \textbf{Algorithm \ref{alg:beamsearch}}) to find the top $k$ items. 30\% ground truths of training queries are provided to train the agent and recall 1@1 is used as the measure metric. The results are presented in \textbf{Figure \ref{fig:vary_training_dcs}}. We can see that the curve will convergence quickly as the training IPC increases for each dataset. By the results of \textbf{Amazon} and \textbf{Echonest}, too large training IPC can even damage the performance slightly.  This results indicate that we don't need to set a too large IPC when the agent collects the training instances. This may be because that too large training IPC can let the agent make more wrong decisions when collecting the training instances. 

\begin{figure}[htbp]
\centering
\subfigure{
\begin{minipage}[c]{0.40\linewidth}
\centering
    \includegraphics[width=1\linewidth]{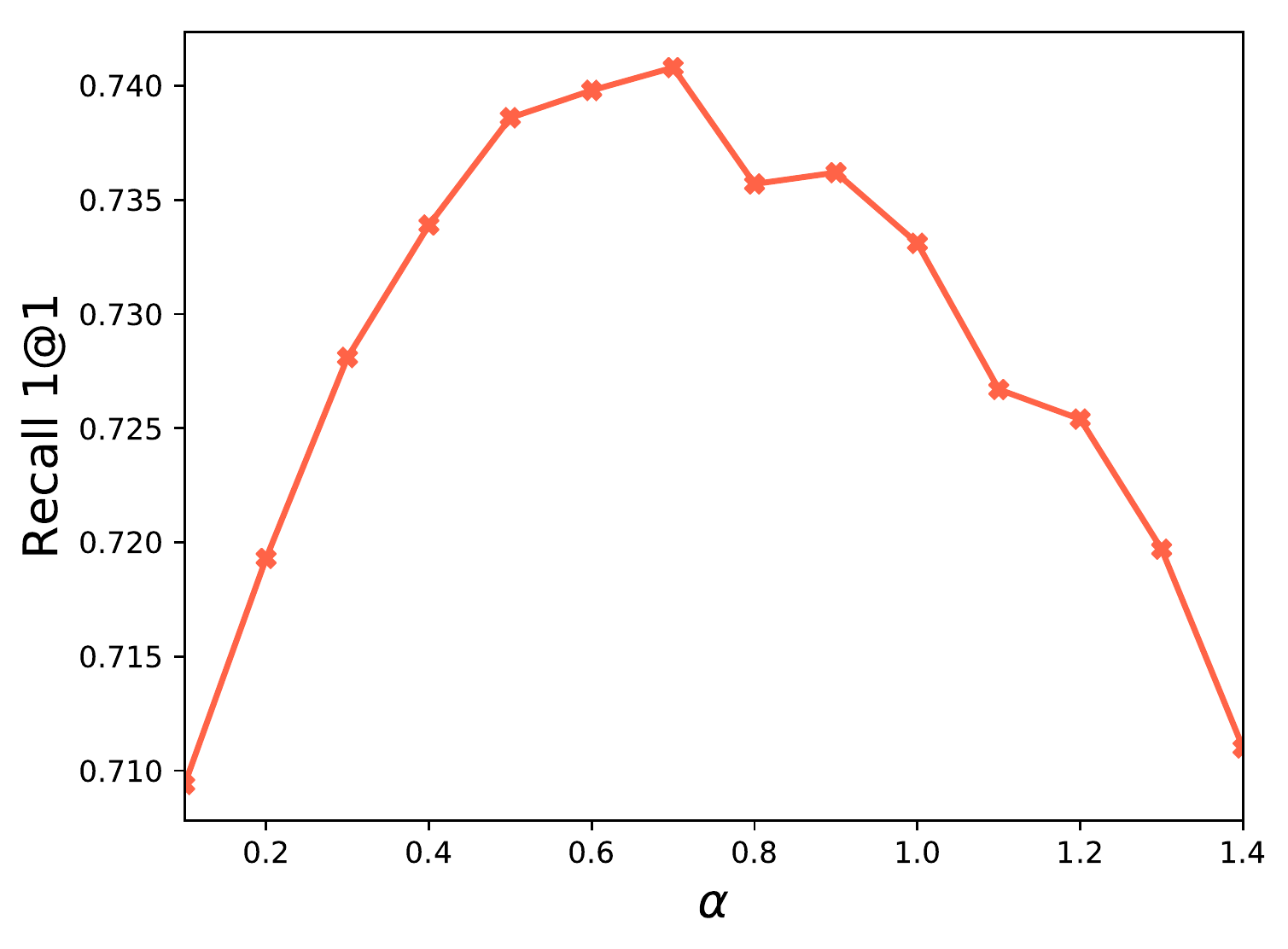}
    (a).Parameter $\alpha$

\end{minipage}}
\subfigure{
\begin{minipage}[c]{0.40\linewidth}
\centering
    \includegraphics[width=1\linewidth]{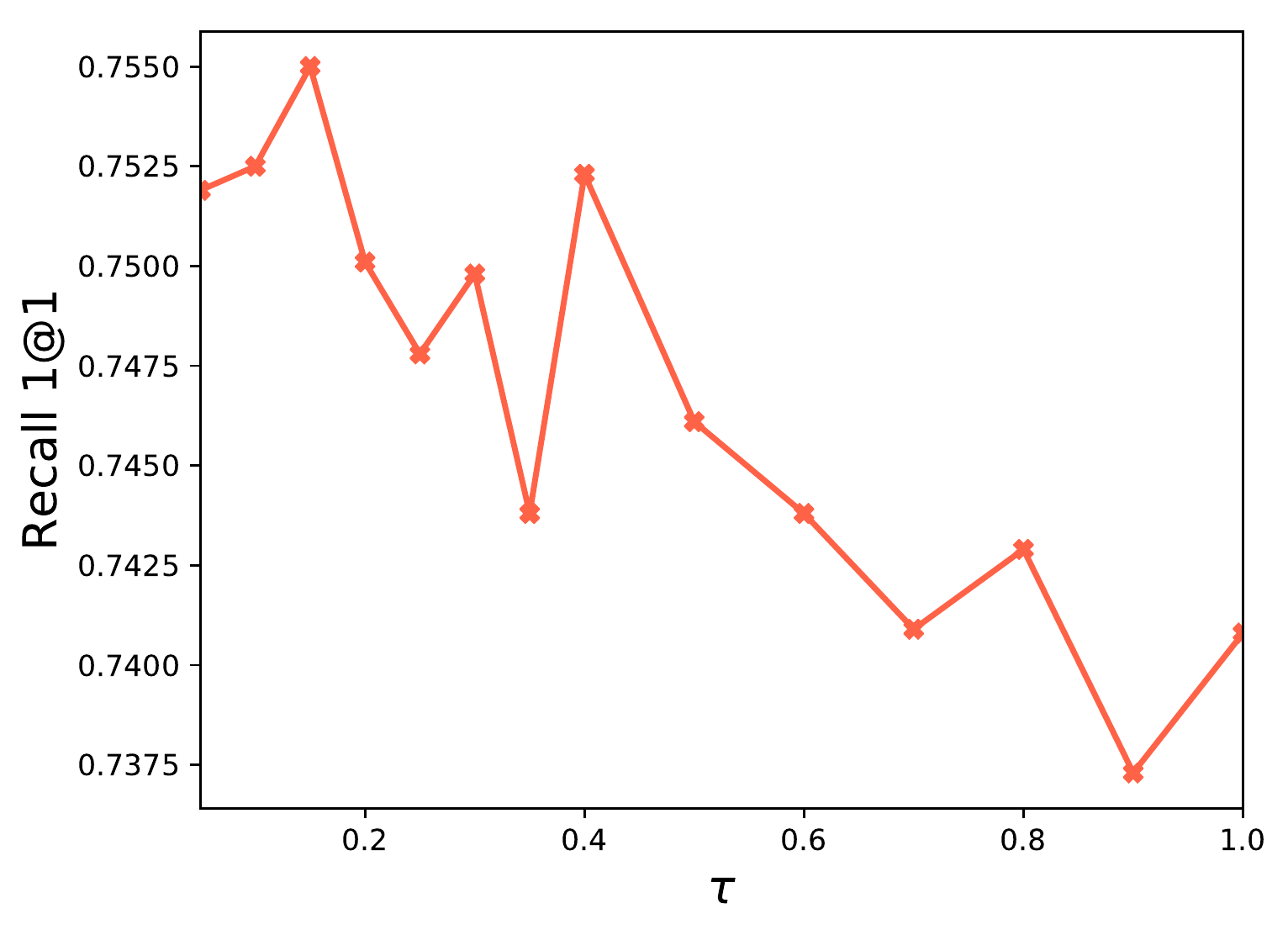}
    (b).Parameter $\tau$
\end{minipage}}
\subfigure{
\begin{minipage}[c]{0.40\linewidth}
\centering
    \includegraphics[width=1\linewidth]{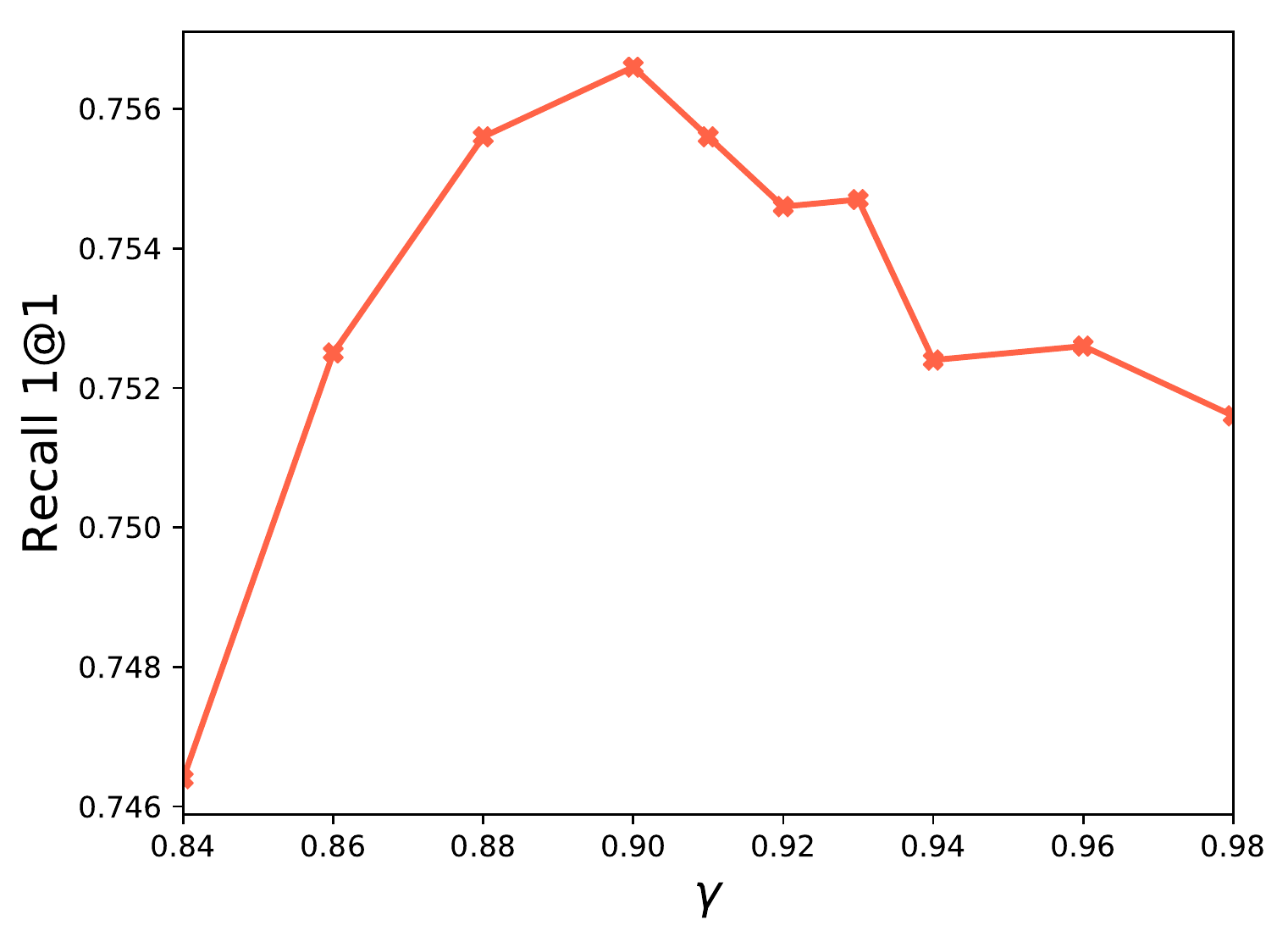}
    (c).Parameter $\gamma$
\end{minipage}}
\subfigure{
\begin{minipage}[c]{0.40\linewidth}
\centering
    \includegraphics[width=1\linewidth]{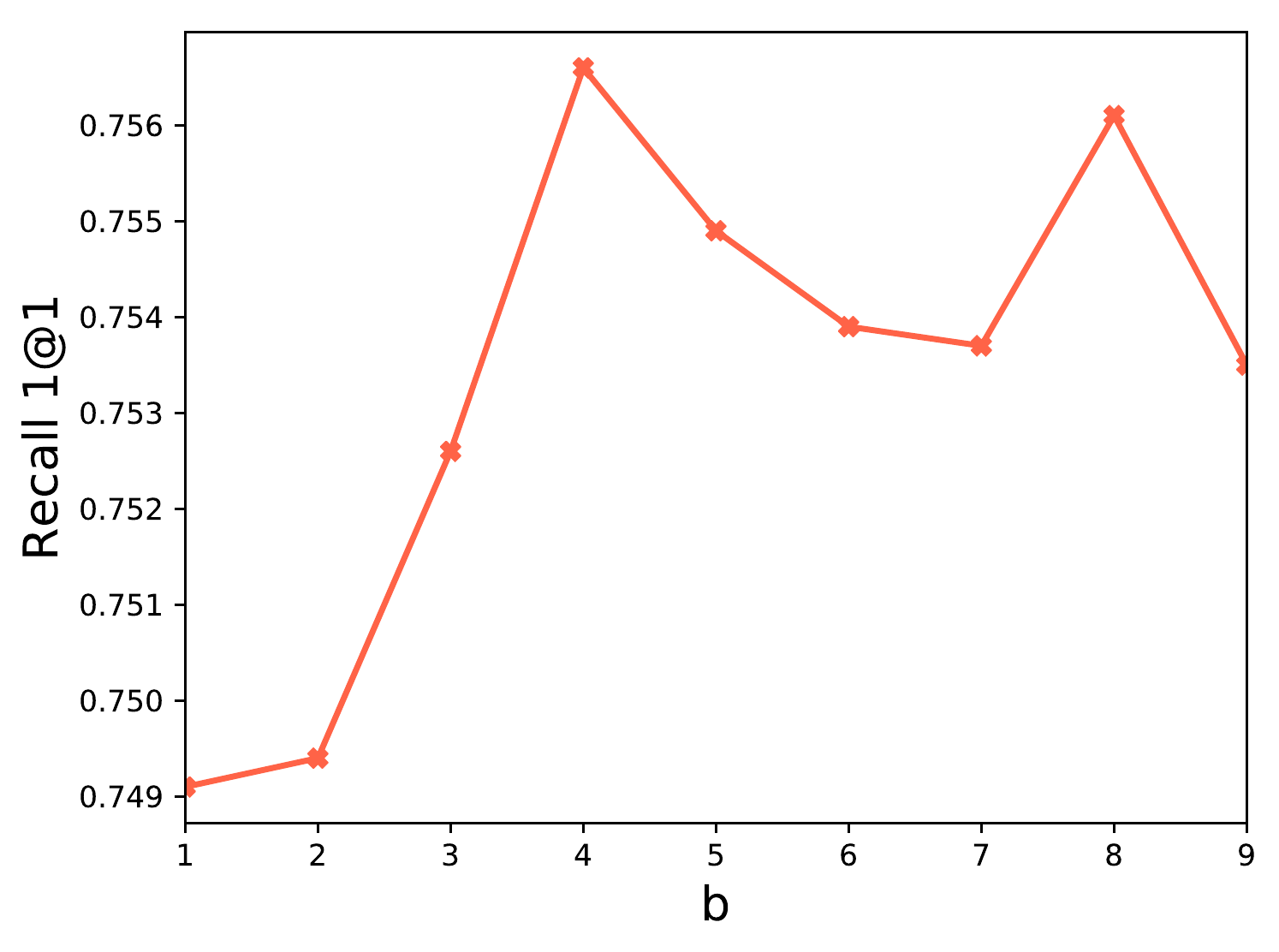}
    (d).Parameter $b$
\end{minipage}}
\caption{The results of varying the hyper-parameters}\label{fig:parameters}
\end{figure}

We also show the influence of the hyper-parameters on ip-NSW graph of \textbf{MovieLens}. The IPC is set to be 256 for both collecting the instances and beam search. 30\% ground truths of the training queries are provided to train the agent. The results are presented in \textbf{Figure \ref{fig:parameters}}. The parameter $\alpha$ is used the trade off the weight of reward shaping if we have some ground truths of the training queries. We fix $\gamma=0.9$, $\tau=1.0$ and $b=4$ when we investigate the influence of $\alpha$. The results of varying the parameter $\alpha$ are presented in the subfigure (a). As the increasing of $\alpha$, the recall 1@1 rises first and then falls. The trend of the curve is in line with our intuition that too small or too large $\alpha$ is not proper. The parameter $\tau$ is temperature that used to control the smoothness of the softmax distribution when the agent selects a candidate from the candidate set. We fix $\alpha=0.7$, $\gamma=0.9$ and $b=4$ when we investigate the influence of $\tau$. The results of varying $\tau$ are presented in subfigure (b). We can see that the recall 1@1 increases at the beginning and then falls in general. The parameter $\gamma$ is the discount factor of the cumulative reward. We fix $\alpha=0.7$, $\tau=0.15$ and $b=4$ when we investigate the influence of $\gamma$. The results of varying $\gamma$ are presented in subfigure (c). The curve nearly has the same trend as subfigure (a). The parameter $b$ is the number of sampled candidates as the baseline of the reward in \textbf{E.q. (\ref{eq:discunt_cumulative_reward})}. We fix $\alpha=0.7$, $\tau=0.15$ and $\gamma=0.9$ when we investigate the influence of $b$. The results of varying $b$ are presented in subfigure (d). The curve increase at the beginning and then falls in general. By \textbf{Figure \ref{fig:parameters}}, we can know that $\alpha$ is the most import parameter and has the most significant influence.

\subsection{Ablation study}
\begin{table}[htbp]
\caption{The results of varying the measure function of NSW.}
\begin{tabular}{|l|l|l|l|l|}
\hline
                                                 &             & ip-NSW & $l_2$-NSW & cos-NSW \\ \hline
\multicolumn{1}{|c|}{\multirow{2}{*}{MovieLens}} & Recall1@1   & $\textbf{0.681}$  & 0.000   & 0.580   \\ \cline{2-5}
\multicolumn{1}{|c|}{}                           & Recall10@10 & $\textbf{0.425}$  & 0.000   & 0.371   \\ \hline
\multirow{2}{*}{Amazon}                          & Recall1@1   & $\textbf{0.823}$  & 0.000   & 0.457   \\ \cline{2-5}
                                                 & Recall10@10 & $\textbf{0.543}$  & 0.000   & 0.329   \\ \hline
\multirow{2}{*}{Echonest}                        & Recall1@1   & $\textbf{0.868}$  & 0.126   & 0.573   \\ \cline{2-5}
                                                 & Recall10@10 & $\textbf{0.547}$  & 0.089   & 0.344   \\ \hline
\end{tabular}
\label{table:ablation_graph}
\end{table}

To investigate the importance of proximity graph for our proposed model, we construct two variants of ip-NSW. Concretely, we still use the procedure of \textbf{Algorithm \ref{alg:nsw}} to construct graph,  but replace the similarity function by negative $l_2$-distance (i.e. $s(\bm{a},\bm{b})=-||\bm{a}-\bm{b}||$)  and cosine similarity (i.e. $s(\bm{a},\bm{b})=\frac{\langle \bm{a},\bm{b}\rangle}{||\bm{a}||\cdot||\bm{b}||}$ ). We call the new graphs as $l_2$-NSW and cos-NSW respectively. IPC is 256 when the agent collects the training instances and  when we test the agent by beam search (i.e. \textbf{Algorithm \ref{alg:beamsearch}}). The top $1$ candidates of ip-NSW are regarded as the approximate ground truths to train our agent on each graph respectively. The 
results are presented in \textbf{Table \ref{table:ablation_graph}}. Our methods perform best on ip-NSW graphs and the advantages are significant. The results on $l_2$-NSW graphs are worst for all cases. These results indicate that the proximity graphs are important for our proposed model. If the proximity graph is not proper, we can't train an good agent for MIPS problem. In addition, the measure function used when constructing the proximity has significant influence for the proximity graph although based on the same construction procedure. 

\begin{table}[b]
\caption{No ground truth vs 30\% ground truths vs approximate ground truths}
\begin{tabular}{|l|l|l|l|}
\hline
          & no gt      &30\% gt & appro gt  \\ \hline
MovieLens & 0.701&\textbf{0.757} & 0.681          \\ \hline
Amazon    & 0.820        & \textbf{0.850} & 0.823 \\ \hline
Echonest  & 0.934&\textbf{0.951} & 0.868          \\ \hline
\end{tabular}
\label{table:gt_approxiamte_gt}
\end{table}
When wen can't obtain all the ground truths for training queries, we can use the approximate ground truths to train the agent. Will the fake ground truths mislead the agent? We summarize the result in \textbf{Table \ref{table:gt_approxiamte_gt}}. The IPC is 256 when the agent collects training instances and the beam search is conducted to search on the proximity graph. Here, we use ip-NSW graph as the proximity graph. The second column (i.e. no gt) means to train the agent with the naive reward function \textbf{E.q. (\ref{reward:no_gt})}. The third column (i.e. 30\% gt) means to train the agent by the queries where 30\% queries have ground truths and left 70\% queries have no ground truths. The fourth column (i.e. appro gt) means to train the agent by the queries with approximate ground truths found by ip-NSW algorithm. The recall 1@1 is reported.
The results on \textbf{Amazon} indicate that the approximate ground truths indeed improve the agent but 30\% ground truths is the better choice. The results on \textbf{MovieLens} and \textbf{Echonest} show that the recall under approximate ground truths is worse than the recall under the no ground truth case. These results indicate that some fake ground truths mislead the agent evidently. Although some approximate ground truths may mislead the agent, our proposed method is still a good choice compared with other algorithms by the aforementioned results. But if the users have extra computation resources, computing more ground truths can be helpful.

\begin{table}[]
\caption{Ablation study to verify the effectiveness of architecture and each part of reward function.}
\begin{tabular}{|l|l|l|l|}
\hline
                           & MovieLens & Amazon & Echonest \\ \hline
\multicolumn{1}{|c|}{Ours} & $\textbf{0.757}$  & $\textbf{0.850}$  & $\textbf{0.951}$    \\ \hline
Ours(shaping)         & 0.194     & 0.273  & 0.408    \\ \hline
Ours(no-baseline)           & 0.267     & 0.255  & 0.167    \\ \hline
\end{tabular}
\label{ablation:architecture}
\end{table}

Lastly, we verify reward functions are effective. Compare proposed algorithm with the following methods.\\
 \textbf{Ours(shaping):} Only remain the reward shaping part of reward function.\\
 \textbf{Ours(no-baseline)} Remove the baseline of \textbf{E.q. (\ref{eq:discunt_cumulative_reward})},\\
IPC=256 when training and testing the agent. 30\% ground truths are given to train the agent on the ip-NSW graph of each dataset. Beam Search (\textbf{Algorithm \ref{alg:beamsearch}}) is applied to find the top $1$ candidate for each query. Recall 1@1 is reported and the results are presented in \textbf{Table \ref{ablation:architecture}}. Our method always performs best. If we only force the agent to follow the shortest paths, i.e. Ours(shaping) algorithm, it can lead to significant bad generalization. Subtracting a baseline from reward function (see \textbf{E.q. (\ref{eq:discunt_cumulative_reward})}) not only makes faster convergence and low variance but also improves the performance of the trained agent.

\section{Conclusion}
In this paper, we introduce a reinforcement learning algorithm to route on proximity graph for MIPS problem. If there is no ground truth for each training query, we can use the queries to train the agent. Besides, with a few ground truths as labels, the agent can utilize these labels to get a better routing policy by learning from these the expert's knowledge. Compared with the existed imitation learning algorithm, our algorithm can utilize a few demonstrations to guide the training process by reward shaping which is more practical. Our empirical studies verify the effectiveness of reward shaping and the superiority of our algorithm compared with the state-of-the-art methods. Once the agent is trained, we can get the latent embedding vectors of all vertices by the GCN of the agent, then our algorithm start an offline search and doesn't need other online cost.
\begin{acks}
The work was supported by grants from the National Key R\&D Program of China under Grant No. 2020AAA0103800, the National Natural Science Foundation of China (No. 61976198 and 62022077) and the Fundamental Research Funds for the Central Universities (No. WK2150110017).
\end{acks}
\bibliographystyle{ACM-Reference-Format}
\bibliography{sample-base}

\end{document}